%% file: TII-15-1127.tex
\documentclass[journal,a4paper]{IEEEtran} 

\usepackage{times}
\usepackage{graphicx}
\usepackage{cite}
\usepackage{units}
\usepackage{amsmath}
\usepackage{color}
\usepackage{enumerate}
\usepackage{anyfontsize}
\usepackage{multirow}
\usepackage{rotating}
\usepackage{url}
\usepackage{mathcomp}
\usepackage{arydshln}
\usepackage{textcomp}

\newcommand{\wred}{\emph{Wi-Red}}
\newcommand{\bwred}{\emph{basic Wi-Red}}
\newcommand{\pow}{PoW}
\newcommand{\std}{DCF}

\newcommand{\rsta}{\mathrm{s}}

\usepackage[english]{babel}
\usepackage{lipsum}% For 'Lorem Ipsum' dummy text
\usepackage[pscoord]{eso-pic}% The zero point of the coordinate systemis the lower left corner of the page (the default).

\newcommand{\placetextbox}[3]{% \placetextbox{<horizontal pos>}{<vertical pos>}{<stuff>}
  \setbox0=\hbox{#3}% Put <stuff> in a box
  \AddToShipoutPictureFG*{% Add <stuff> to current page foreground
    \put(\LenToUnit{#1\paperwidth},\LenToUnit{#2\paperheight}){\vtop{{\null}\makebox[0pt][c]{#3}}}%
  }%
}%

\title{Seamless Link-level Redundancy to Improve Reliability of Industrial Wi-Fi Networks
\thanks{This work was partially supported by the Ministry of Education, University, and Research of Italy (MIUR) in the framework of the Project Wi-Fact ``WIreless FACTory and beyond'' (DM53543). Copyright (c) 2016 IEEE. Personal use of this material is permitted. However, permission to use this material for any other purposes must be obtained from the IEEE by sending a request to pubs-permissions@ieee.org. The authors are with the National Research Council of Italy, Istituto di Elettronica e di Ingegneria dell'Informazione e delle Telecomunicazioni (CNR-IEIIT), I-10129 Torino, Italy (e-mail: gianluca.cena@ieiit.cnr.it, stefano.scanzio@ieiit.cnr.it, adriano.valenzano@ieiit.cnr.it).}}

\author{Gianluca~Cena, \textit{Senior Member}, \textit{IEEE}, Stefano~Scanzio, \textit{Member}, \textit{IEEE}, and\\Adriano~Valenzano, \textit{Senior Member}, \textit{IEEE}}

\begin{document}

\placetextbox{0.5}{1}{This is the author's version of an article that has been published in this journal.}
\placetextbox{0.5}{0.985}{Changes were made to this version by the publisher prior to publication.}
\placetextbox{0.5}{0.97}{The final version of record is available at \href{https://doi.org/10.1109/TII.2016.2522768}{https://doi.org/10.1109/TII.2016.2522768}}%
\placetextbox{0.5}{0.05}{Copyright (c) 2016 IEEE. Personal use is permitted.}
\placetextbox{0.5}{0.035}{For any other purposes, permission must be obtained from the IEEE by emailing pubs-permissions@ieee.org.}%
\placetextbox{0.5}{0.1}{(Winner of the ``\textbf{2017 Best Paper Award for the IEEE Transactions on Industrial Informatics}'')}

\maketitle

\begin{abstract}
The adoption of wireless communications and, in particular, Wi-Fi, at the lowest level of the factory automation hierarchy has not increased as fast as expected so far, mainly because of serious issues concerning determinism. Actually, besides the random access scheme, disturbance and interference prevent reliable communication over the air and, as a matter of fact, make wireless networks unable to support distributed real-time control applications properly. Several papers recently appeared in the literature suggest that diversity could be leveraged to overcome this limitation effectively.

In this paper a reference architecture is introduced, which describes how seamless link-level redundancy can be applied to Wi-Fi.
The framework is general enough to serve as a basis for future protocol enhancements, and also includes two optimizations aimed at improving the quality of wireless communication by avoiding unnecessary replicated transmissions.
Some relevant solutions have been analyzed by means of a thorough simulation campaign, in order to highlight their benefits when compared to conventional Wi-Fi.
Results show that both packet losses and network latencies improve noticeably.
\end{abstract}

%------------------------------------------------------------------------- 

\section{Introduction}
Wireless communication technologies, and particularly industrial wireless sensor networks (IWSNs) \cite{2014-TII-Guest}, are used more and more in automation scenarios, including real-time control systems \cite{2014-TII-Proto}. Two of the most popular solutions for process automation at the time of writing are WirelessHART and ISA100.11a \cite{2011-IEM-Petersen}. Although they basically rely on IEEE 802.15.4 \cite{2011-std-802154} for frame transmission over the air, mechanisms like channel hopping and channel blacklisting have been included in order to make communication more resilient to interference and disturbance, which often abound in industrial environments \cite{2014-TII-inteferences}.
{Instead, for factory automation (which has tighter timing constraints), commercial solutions like the wireless interface for sensors and actuators (WISA) and the industrial wireless LAN (IWLAN) were developed by ABB and Siemens, respectively.
The former is based on IEEE 802.15.1 (Bluetooth), enhanced by a specific full-duplex access scheme, whereas the latter employs an extension to IEEE 802.11 \cite{2012-std-80211}, which exploits one or more centralized coordinators and foresees optimized multiple antennas or radiating cables to ensure reliable communication.

With the exception of IWLANs, IEEE 802.11 networks are usually deemed not reliable enough for real-time control applications \cite{2013-IEM-Vitturi}.
Although \emph{rate adaptation} techniques \cite{2013-TII-Vitturi-Rate}, customarily employed in commercial Wi-Fi equipment, allow tolerating a decrease in the signal-to-noise ratio (SNR) by switching to more robust modulations schemes based on lower transmission rates, they are able to mitigate the instability of wireless communication only in part.

The adoption of techniques like channel hopping and blacklisting in Wi-Fi is probably not a viable option either.
Although a specific physical layer based on frequency hopping spread spectrum (FHSS) was foreseen since the initial IEEE 802.11 specification, it never becomes very popular and recently it was declared obsolete.
In theory, a mechanism could be layered on top of orthogonal frequency-division multiplexing (OFDM) so that every wireless station (STA) synchronizes to beacons and hops---at faster pace---between channels.
However, association of conventional STAs would be precluded, which poses coexistence issues with existing devices.

Ensuring high availability in industrial automation systems that include wireless links can be accomplished by exploiting \emph{diversity} \cite{2014-TII-Diversity}.
For example, in the case of Wi-Fi, simultaneous dual-band equipment such as access points (APs) and wireless bridges (WBs) can be used to this purpose. 
In the case a problem is experienced that blocks communication over a channel for a protracted period of time, the spanning tree protocol (STP) or its rapid variant (RSTP) take care of dynamically selecting a new path, which includes wireless links operating at a different frequency.
Unfortunately, recovery times for these solutions can be as long as several seconds \cite{2007-ETFA-Prytz}, which makes them unsuitable for the typical real-time applications found at the shop-floor level.

The extended channel switching (ECS) procedure defined in IEEE 802.11 allows to deal with situations where disturbance affecting the wireless channel grows excessively.
This is obtained by letting the AP notify the STAs that the channel frequency is going to be changed.
Although the transition is quite fast in this case, frames may be lost (or delayed unacceptably) before the decision to switch channel is made.

A different approach relies on a particular modulation and coding scheme (MCS), known as MCS 32, included in the IEEE 802.11n specification.
This technique exploits channel bonding by sending the same frame at $\unit[6]{Mbps}$ on both halves of a single $\unit[40]{MHz}$ channel. MCS 32 is noticeably more robust than other schemes, but frame transmissions require that both the bonded channels be available at the same time, and this may reduce network utilization and increase latency.

\emph{Seamless redundancy}---as specified in  IEC 62439-3 \cite{2012-std-PRP}, also known as parallel redundancy protocol (PRP)---is probably the best approach to ensure both timeliness and high dependability in control systems based on real-time Ethernet (RTE) technology.
In PRP, each frame is sent by transmitters on two distinct networks.
The two copies reach their destination(s) through different paths and, possibly, at different times.
The first copy that arrives to destination is retained and delivered to the application (or the upper protocol layers), whereas the second one (i.e., the \emph{duplicate}) is discarded.
This solution has two advantages: first, frames are lost only when they are dropped on both networks. This means that the likelihood that a packet is not delivered because of faults affecting network equipment and media can be lower (even noticeably lower) than for non-redundant solutions. Second, the transmission latency decreases as well, and corresponds to the minimum among the times taken by all the different copies of each frame to reach the destination through the relevant networks paths.

PRP is meant to both provide seamless \emph{end-to-end} communication redundancy and grant a certain degree of fault tolerance to critical applications in highly-dependable systems.
In \cite{2012-WFCS-WoP1} it was used, together with conventional Wi-Fi equipment, to increase the communication reliability of a wireless link between two nodes.
In the following, such a solution will be denoted \emph{PRP over Wi-Fi} (\pow{}).
When coupled with proper medium access control (MAC) overlays \cite{2007-TII-vas},\cite{2011-LCN-vas},\cite{2011-ISPCS-TDMA-flexWARE}, \cite{2015-TII-RBIS}, \pow{} can enable quasi-deterministic behavior.

Unfortunately, this approach also exhibits a non-negligible drawback, originating from the transmission of every frame on both channels.
This implies that the network load on each single channel is exactly the same as for non-redundant solutions.
As a result, the overall traffic over the air is doubled, which could be a nuisance as bandwidth in unlicensed portions of the radio spectrum is a scarce resource.

Since the quality of communication may vary suddenly and unexpectedly on wireless channels, packet queues in intermediate relay devices (e.g., APs) could grow considerably. 
As a consequence, latency increases and devices may even run out of memory, hence causing frames to be dropped.
In such conditions, preventing unnecessary transmissions over the redundant channels is the key to improve the overall network behavior.
Incidentally, doing so also reduces power consumption, which can be quite appealing in mobile devices.

In the following, the \emph{Wi-Fi Redundancy} (\mbox{\wred{}}) solution is presented, which is aimed at providing seamless \emph{link-level}  redundancy in IEEE 802.11 networks.
It also includes mechanisms that make it less bandwidth-hungry than PoW, by reducing the amount of duplicate frames sent on air.
Unlike approaches like \cite{2014-WFCS-Mifdaoui}, \wred{} relies on proven \mbox{Wi-Fi} technology, which makes it a natural solution to deploy wireless extensions of Ethernet, especially in industrial plants.
Moreover, \wred{} is conceived to grant perfect coexistence with conventional Wi-Fi.

The basic principles behind \wred{} were first introduced in \cite{2014-WFCS-WiRed} and \cite{2014-ETFA-DDD}.
With respect to those preliminary papers, several relevant aspects have been taken into account here, including in particular the extension of redundancy mechanisms to multiple channels, their relationships with Quality of Service (QoS), 
and their effects on the order of received frames.
Instead, frame aggregation and block acknowledgment are not considered in this paper, since possible benefits due to the inclusion of optimizations in redundant transmission are still questionable in these cases and need further investigation.
Moreover, such mechanisms are useful when most frames spend  non-negligible amounts of time queued for transmission, but this is unlikely to occur in the case of process data exchanged in typical real-time industrial applications.

The resulting reference architecture is suitable for describing the existing proposals and general enough to provide a solid basis for future enhancements, including the definition of additional heuristics and cross-layer integration with upper protocols (e.g., the MAC) in order to improve determinism further.

The paper is structured as follows: in Section \ref{sec:WiRed} a general framework is introduced to deal with seamless redundancy in Wi-Fi, while in Section \ref{sec:RedProt} the \wred{} protocol is presented with focus on two mechanisms that prevent transmissions of useless duplicates.
Section \ref{sec:PerfEval} describes the testbed we used for experimental evaluation.
Finally, Section \ref{sec:Results} reports on the results of some simulation campaigns and performs a quick comparison of the different techniques.
\begin{figure}
  \centering
  \includegraphics[width=\columnwidth]{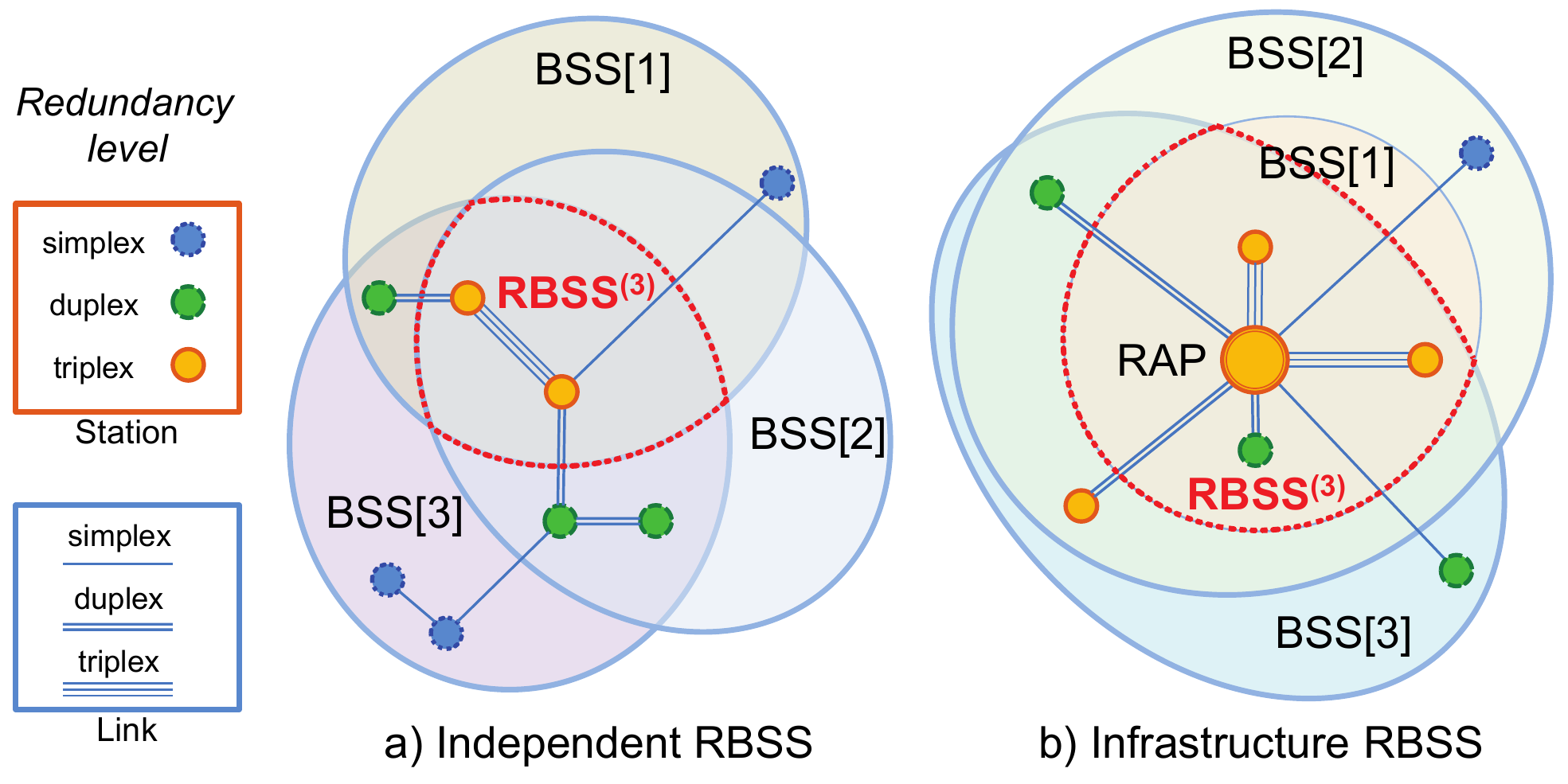}
  \caption{RBSS and redundancy levels (station/link).}
  \label{fig:rbss}
\end{figure}

\section{Wi-Fi Redundancy Framework}\label{sec:WiRed}
Basically, \wred{} is seamless redundancy---as defined by PRP---applied to Wi-Fi.
Unlike PRP, which was conceived as an end-to-end solution to provide fault tolerance in Ethernet-based systems, \wred{} operates on each single wireless link---either between a STA and the AP in an infrastructure basic service set (BSS) or between two STAs in an independent basic service set (IBSS).
It aims at enhancing the quality of communication in Wi-Fi to a degree that, in theory, could resemble wired connections.

Another difference between PRP and \wred{} is the way duplicate copies of the same frame are detected by receivers.
In PRP a redundancy control trailer (RCT) is appended to every Ethernet frame for its unambiguous identification by end nodes.
This is not needed in \wred{}, at least in theory, since the \emph{sequence control} field included in each IEEE 802.11 MAC protocol data unit (MPDU) can be used to this extent.
In practice, keeping sequence numbers aligned across different MACs operating on distinct channels is hardly feasible with existing chipsets.
Therefore, software implementations of \wred{} may still make use of RCT. 
This can be obtained quite easily, since the maximum transmission unit (MTU) in IEEE 802.11 is larger than the Ethernet MTU.

\subsection{Network Architecture}
As shown in Fig.~\ref{fig:rbss}, a \wred{} network consists of a number of \emph{redundant stations} (RSTAs), each of which is joined to two (or more) distinct BSSs.
Generically speaking, the intersection of these BSSs is referred to as \emph{redundant BSS} (RBSS) and includes all the related RSTAs.
In the following text, $\rsta{}_m$ is the $m$-th RSTA in the considered RBSS, while each BSS in the RBSS is denoted BSS[$k$], $k \in [1...R_{\operatorname{N}}]$, where $R_{\operatorname{N}}$ is the \emph{network redundancy level} (simplex, duplex, triplex, and so on).

BSSs overlap in space, but operate on different channels or, even better, distinct frequency bands (e.g., $2.4$, $5$, and $\unit[60]{GHz}$).
Besides RSTAs, each BSS may include conventional, non-redundant STAs as well.
The RBSS coverage area comprises at least the places where RSTAs are located, and corresponds to the region where full-redundant communication is allowed.
Non-redundant (or partially-redundant) communication between RSTAs may be possible in areas larger than the RBSS.
\subsection{Redundant Stations}
\begin{figure}
  \centering
  \includegraphics[width=\columnwidth]{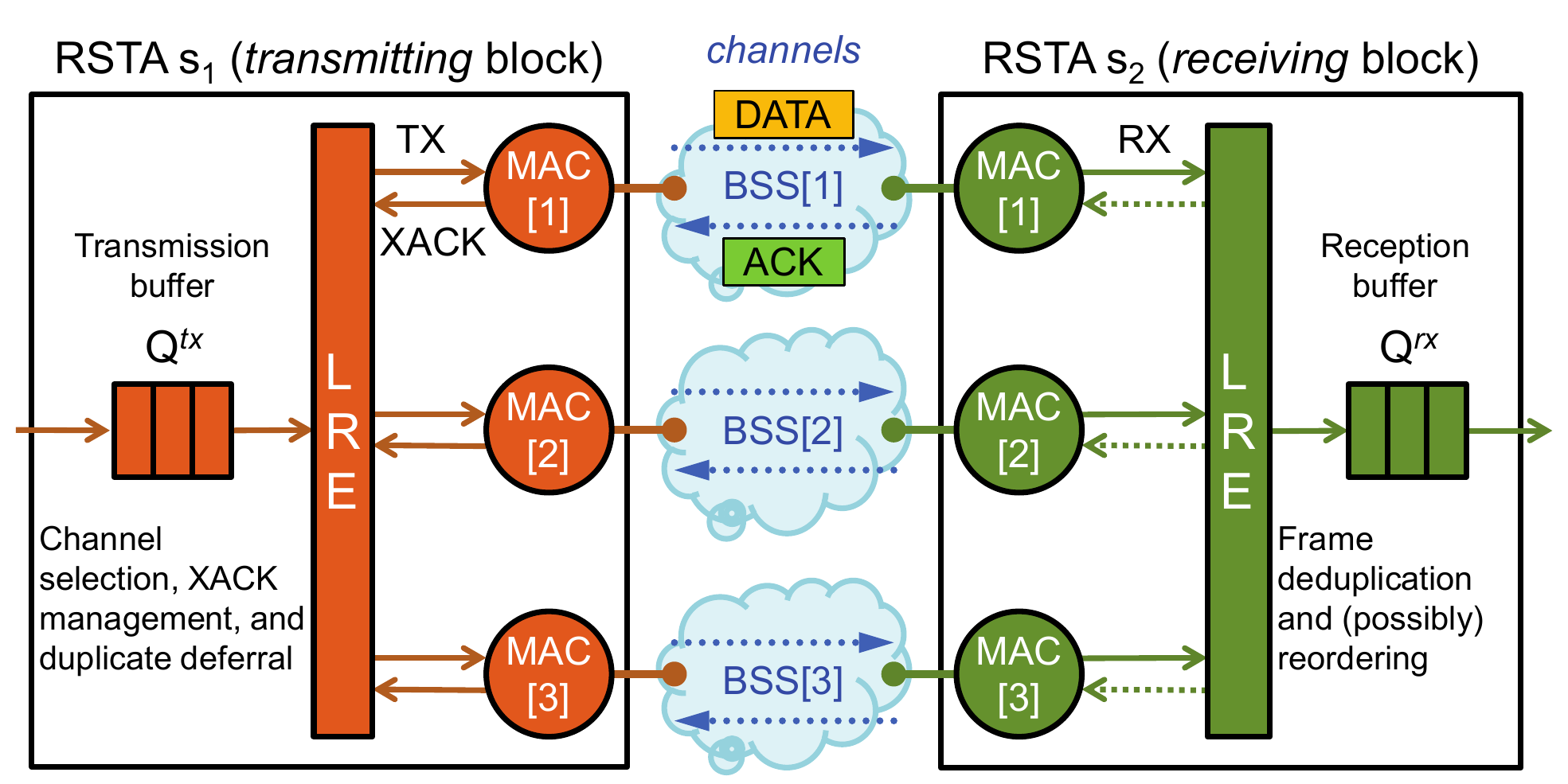}
  \caption{RSTA conceptual block diagram (triplex redundancy).}
  \label{fig:rsta}
\end{figure}
As depicted in Fig.~\ref{fig:rsta}, an RSTA includes several independent entities, or \emph{substations} (sub-STAs), each one being a member of one BSS in the RBSS.
Sub-STAs of RSTA $\rsta{}_m$ are referred to as $\rsta{}_m$[$k$], $k \in [1...R_{\rsta{}_m}]$, where $R_{\rsta{}_m}$ is the \emph{station redundancy level} of $\rsta{}_m$ (i.e., the number of channels simultaneously supported by $\rsta{}_m$).
Each sub-STA implements the physical (PHY) and MAC protocol layers.
Although both layers have to comply to IEEE 802.11 and full compatibility must be retained with Wi-Fi equipment, slight changes might be needed to the MAC in order to enhance performance.

Operations of the different sub-STAs in the same RSTA are co-ordered by a \emph{link redundancy entity} (LRE).
On the transmitter side, LRE takes care of dispatching frames, queued in the \emph{transmission buffer} ($\operatorname{Q}^\mathit{tx}$), to the MACs of the local sub-STAs.
Suitable mechanisms should be introduced in order to prevent useless transmission of duplicates.
RSTA implementations can make use of either a single buffer $\operatorname{Q}^\mathit{tx}$, shared among all the sub-STAs, or separate transmission buffers $\operatorname{Q}^\mathit{tx}[k]$ for different sub-STAs.
In the first case each buffer entry has to include also information about the transmission state of the relevant frame on every BSS (e.g., waiting, transmitting, retransmitting, aborting, transmitted, or discarded).
On the receiver side, LRE  removes duplicates from the \emph{reception buffer} ($\operatorname{Q}^\mathit{rx}$) and, possibly, reorders frames before they are delivered to the upper layers.

\subsection{Redundant Networks}
Real RBSSs are likely to be deployed as infrastructure networks.
Infrastructure RBSSs are based on \emph{redundant access points} (RAPs). 
The RAP (here denoted $\rsta{}_0$) relays frames between either different RSTAs or any RSTA and the distribution system (DS), i.e., the Ethernet backbone.
In this case, $R_{\operatorname{N}}$ is the same as the station redundancy level $R_{\rsta{}_0}$ of the RAP.
Definitions are slightly different for redundant IBSSs, as communications take place between STAs directly.
As this kind of networks is less significant from a practical point of view, it will not be considered in this paper.

The behavior of RSTAs, as observed on any BSS, is not noticeably different from conventional STAs.
This means that each RSTA has to join each BSS in the RBSS separately.
To enable this, the conventional Wi-Fi procedures for AP association must be extended, so as to allow each RSTA to discover all the redundant channels in the RBSS. 
For instance, this information can be included in beacons sent by the RAP and other management frames (association, probe, and so on), by extending their basic format. 
Doing so does not impair backward compatibility with legacy STAs.

To achieve \emph{full redundancy}, every RSTA must be member (through distinct sub-STAs) of all the BSSs in the RBSS.
This is possible provided that: a) all $R_{\operatorname{s}_m}$ values are the same as $R_{\operatorname{N}}$, and b) the actual \emph{link redundancy level} $R_{\operatorname{l}_m}$ of every RSTA $\rsta{}_m$ coincides with $R_{\operatorname{s}_m}$ (i.e., $\rsta{}_m$ is within the RBSS coverage area and all its sub-STAs can communicate on the related BSSs).
In other words, $R_{\operatorname{l}_m}=R_{\rsta{}_m}=R_{\operatorname{N}},~\forall~\rsta{}_m\in$~RBSS.

Conversely, $R_{\operatorname{l}_m}$ is not requested to  be the same as $R_{\operatorname{N}}$ for all RSTAs in order to support \emph{partial redundancy}.
For example, a duplex RSTA may associate with a triplex RAP, and $R_{\operatorname{l}_m}$ is only duplex in this case.
The same happens when a triplex RSTA is located in a place where only two BSSs of a triplex RAP can be reached.
In general $R_{\operatorname{l}_m} \leq \min(R_{\rsta{}_m}, R_{\rsta{}_0})$, and inequality may hold when $\rsta{}_m$ is outside the coverage area of some BSS in the RBSS.
Let RBSS$^{(r)}$ denote an $r-$level RBSS ($r\leq R_{\operatorname{N}}$), defined as the set of all the RSTAs that are enabled to communicate over links whose redundancy level is at least $r$, i.e., RBSS$^{(r)}=\left\{\rsta{}_m |  R_{\operatorname{l}_m}\geq~r\right\}$.
For these sets the condition $r_1<r_2\Rightarrow$~RBSS$^{(r_1)}\supseteq$~RBSS$^{(r_2)}$ holds.

Details about setting up a redundant link and negotiating its minimum allowed redundancy level are outside the scope of this paper.
For the sake of simplicity and with no loss of generality, in the following we will assume full redundancy.

\section{Redundancy Protocol}\label{sec:RedProt}
In \wred{}, a \emph{source} RSTA can send multiple copies of the same frame on different channels of a redundant link.
Duplicates are then detected and discarded by the \emph{target} RSTA(s), as it happens in PRP.
However, in \wred{} these operations are performed directly by the data-link layer (DL).

In the \emph{basic} version of \wred{}, distinct copies of every frame are concurrently sent on all channels.
In this way, the traffic generated by an RSTA on each BSS is the same as a conventional, non-redundant STA in Wi-Fi.
When the overall behavior is considered, this approach closely resembles the \pow{} solution in \cite{2012-WFCS-WoP1}---even though \pow{} actually operates above the DL layer.
As pointed out in \cite{2014-WFCS-WiRed} and \cite{2014-ETFA-DDD}, bandwidth may be wasted and this is not a good option, since the wireless spectrum in unlicensed bands is already crammed.

Two mechanisms were introduced to cope with this drawback, which operate on the transmitter side and are implemented in the LRE.
They are aimed at decreasing the overall traffic on air by preventing  the transmission of inessential duplicates, while improving the quality of communication at the same time.
The first mechanism \emph{reactively} avoids the transmission of duplicates for frames that have already been delivered to the intended receiver.
This conceptually simple approach does not exhibit any drawbacks and always brings benefits.
The second mechanism, instead, works \emph{proactively}, by deferring the transmission of duplicates for frames that are already in the process of being sent.
However, while reducing network traffic, deferral may increase latency.
Hence, it has to be managed carefully.

Although the adoption of these \emph{duplicate avoidance} (DA) mechanisms is highly recommended in \wred{}, their implementation is not mandatory.
Correct communications shall be granted in the RBSS nevertheless, irrespective of the fact that any of the involved RSTAs be actually able to provide such a kind of support.
Both DA mechanisms apply to unicast (\emph{acknowledged}) transmissions only, while no optimization is possible for multicast (\emph{unacknowledged}) communications.
Consequently, multicast frames have to be sent on all channels according to the \bwred{} rules.

Unlike MCS 32, \wred{} exploits \emph{time} diversity besides \emph{frequency} diversity.
In fact, unless wireless channels remain idle for most time (that is quite a favorable condition), different copies of the same frame are typically transmitted at different instants.
This means that \wred{} may bring benefits for both narrow-band disturbance and short wide-band noise pulses.

In the following, we will take into account the behavior of a single unidirectional redundant link, but this is not a limiting assumption. Since acknowledged transmissions are dealt with by the MAC on a per-link basis, the \wred{} redundancy scheme and the related optimizations are independently applied to each communication between RSTA pairs and, in particular, to each \emph{uplink} and \emph{downlink} between the RAP and an RSTA.
Overall, each frame received by the RAP over any redundant uplink undergoes duplicate detection and removal before forwarding. 
Moreover, as soon as a valid copy of a frame addressed to an RSTA becomes available in the RAP (this holds also for frames received on Ethernet ports), it is queued for transmission on the relevant redundant downlink.

\subsection{Reactive Duplicate Avoidance}
Transmissions of redundant copies  \emph{after} a frame has already been delivered successfully on another channel are completely useless and should be avoided.
The sender can infer the occurrence of such a situation by exploiting the \emph{acknowledgment} (ACK) mechanism, which is used in the distributed coordination function (DCF) of IEEE 802.11 to confirm the correct delivery of unicast data frames (DCF is the most basic MAC scheme in Wi-Fi, on which all the other coordination functions rely).
DA approaches based on this principle are referred to as \emph{reactive duplicate avoidance} (RDA).
According to DCF, every time a STA transmits a frame, a timeout (\emph{ACKTimeout}) is started.
If an ACK frame is not received from the intended destination before \emph{ACKTimeout} expires, a \emph{retry counter} ($\operatorname{RC}$) is increased and a retransmission is attempted.
When $\operatorname{RC}$ reaches a predefined \emph{retry limit} ($\operatorname{RL}$), retransmissions end and the frame is discarded.
\begin{table*}
  \caption{Main features of Duplicate Avoidance mechanisms to reduce network load in \wred{} -- full support ($\bullet$) / partial support ($\circ$).}
  \label{tab:TbD}
  \begin{center}
    \footnotesize
    \begin{tabular}{l||c|c|c|c|c|c}
      Operation	& RDA/Q & RDA/R & RDA/M & RDA/Q+PDA & RDA/R+PDA & RDA/M+PDA \\
      \hline \hline
      Removes duplicates from transmission queues upon XACKs	     & $\bullet$ & $\bullet$ & $\bullet$ & $\bullet$ & $\bullet$ & $\bullet$ \\ 
      Aborts ongoing MAC transmissions of duplicates upon XACKs	     &           & $\circ$   & $\bullet$ &           &  $\circ$  & $\bullet$ \\
      Defers further MAC transmissions of duplicates upon heuristics &           &           &           & $\bullet$ & $\bullet$ & $\bullet$ \\
      \hline
    \end{tabular}
    \end{center}
\end{table*}
\wred{} optimizes transmissions by means of \emph{cross acknowledge} (XACK) events. An XACK is generated inside RSTA $\rsta{}_m$ whenever one of its sub-STAs receives an ACK frame for a pending frame exchange.
Such an event can typically be inferred from the interrupts raised by wireless adapters at the end of each frame transmission.
Every XACK is tagged with the channel \emph{k} where the ACK was received (that is, BSS[$k$]) and a unique identifier \emph{i} of the relevant data frame (F$_i$), which can be easily derived from the MAC sequence control field.
On reception of the XACK(\emph{k},~\emph{i}) event, every sub-STA $\rsta{}_m[g]$ in the RSTA, $g \neq k$ (that is, with the exception of $\rsta{}_m[k]$), searches its transmission queue $\operatorname{Q}_m^\mathit{tx}[g]$ for a corresponding data frame and, if a match occurs, the matching entry is removed.
If the RSTA adopts a single transmission buffer $\operatorname{Q}_m^\mathit{tx}$, then LRE is in charge of removing useless duplicates.
We call the above mechanism RDA/Q: on most platforms, it can be implemented by introducing changes in the device drivers of wireless adapters.

RDA/Q can be improved by allowing sub-STAs to abort ongoing transmissions, too.
To do so, when XACK(\emph{k},~\emph{i}) is triggered by $\rsta{}_m[k]$, the MACs of the other sub-STAs $\rsta{}_m[g]$ in the RSTA are inspected to determine if frame F$_i$ is currently in the process to be sent. 
When this occurs, the transmission (or retransmission) of F$_i$ on $\rsta{}_m[g]$ is aborted. 
A suitable \emph{abortable} DCF (ADCF) technique was described in \cite{2014-WFCS-WiRed}, which requires MAC modifications but retains full compatibility with DCF.
It can be proven that frame transmissions can be safely aborted (that is, without impairing coexistence with legacy Wi-Fi STAs) only between retries.
In practice, when an abort request is issued in a sub-STA, its MAC has to wait for the end of the current transmission attempt (including the ACK reception).
If the ACK is received before \emph{ACKTimeout} expiration, abortion is unnecessary.
Otherwise, the MAC keeps on waiting until \emph{ACKTimeout} expires, and only then it cancels the transmission and resets its \emph{contention window} (CW) as if the frame were discarded.
More details about ADCF can be found in \cite{2014-WFCS-WiRed}.
We call RDA/M the adoption of the above MAC-level transmission abort mechanism in conjunction with RDA.

Deep changes are needed to the DCF protocol state machines to implement RDA/M. 
Instead, a much simpler solution can be devised, we call RDA/R, which is based on a streamlined version of ADCF.
In this case, in order to abort an ongoing frame transmission in the MAC of $\rsta{}_m[g]$, its retry counter $\operatorname{RC}_m[g]$ is set equal to the retry limit $\operatorname{RL}$, as suggested in \cite{2010-CPSCom-802.11_dinamic_tuning}.
While RDA/R requires limited changes to DCF, this arrangement is sub-optimal since the check on $\operatorname{RC}$ for any given frame is carried out \emph{after} each transmission attempt.
This means that frames, which were already fed into the MAC and whose transmission has still to begin at the instant when the abort request is issued, nevertheless undergo a transmission attempt.
Preliminary experiments, carried out on a partial implementation of RDA/M, showed that, although it is much more complex than RDA/R, it only has a slight performance edge.
Because of its intrinsic simplicity, RDA/R is a satisfactory compromise between benefits and complexity.
For this reason it has been considered in the following analysis in the place of RDA/M.

Summing up, upon XACK events RDA/Q is only able to remove duplicate frames queued in the transmission buffers of the sub-STAs, whereas RDA/R (and RDA/M) are also provided with the ability of aborting ongoing MAC transmissions.
Conversely, XACKs are not dealt at all by \bwred{}, which transmits every single frame on all channels.
Because of their different ability to prevent duplicate transmissions, we must expect increasing performance when moving from \bwred{} to RDA/Q and from RDA/Q to RDA/R.

\subsection{Proactive Duplicate Avoidance}
A second class of DA mechanisms for \wred{} is meant to \emph{complement} and improve RDA.
They spring into action whenever the MAC of a sub-STA concludes a frame transfer and a new frame has to be selected for transmission from the RSTA queue (provided that at least one frame is present there).
Let $\rsta{}_m[k]$ be that sub-STA and F$_h$ the frame in the first position of its queue (i.e., the oldest one), $\operatorname{F}_h=\operatorname{head}(\operatorname{Q}_m^\mathit{tx}[k])$.
According to DCF, $\operatorname{F}_h$ should be immediately fed into the MAC of $\rsta{}_m[k]$ for transmission.
Performance can be improved by checking whether F$_h$ is already being sent by other sub-STAs in $\rsta{}_m$.
In this case, the transmission of a duplicate of F$_h$ by $\rsta{}_m[k]$ can be temporarily deferred and another pending frame in $\operatorname{Q}_m^\mathit{tx}[k]$ can be selected and sent in the place of F$_h$.

We call this kind of mechanism \emph{proactive duplicate avoidance} (PDA), as it aims at preventing transmissions that are likely unnecessary, in particular when disturbance and interference are at acceptable levels.
In fact, if an XACK is received for a frame whose redundant copy has been deferred, a duplicate (and hence, useless) transmission is avoided.
The higher the number of redundant channels, the larger the benefits that PDA techniques achieve.
However, special attention must be paid when dealing with time-critical communications, in which case the effect of deferral on transmission latency has to be carefully evaluated.

Using PDA without RDA, although in theory possible, is mostly pointless in practice.
By disabling RDA, in fact, a frame would no longer be removed from all the sub-STAs in the sender following the reception of the related ACK on any channel.
Because of deferrals, the frame transmission order may vary across channels, but the overall amount of network traffic remains the same as in \bwred{}.
As a consequence, no noticeable improvements have to be expected from this arrangement over the much simpler \pow{}.
For this reason, PDA should be regarded uniquely as a set of heuristics aimed at improving RDA behavior, by reducing further the likelihood that duplicate copies are sent over the air.

Different approaches can be adopted to decide whether a duplicate transmission has to be deferred. A very simple solution for duplex RSTAs is the \emph{dynamic duplicate deferral} (DDD) technique \cite{2014-ETFA-DDD}.
DDD works as follows: the MAC in the sub-STA of $\rsta{}_m$ that works on channel $\overline{k}\not= k$, i.e., $\rsta{}_m[\overline{k}]$, is checked to know if it is in the process of sending F$_h$. If the answer is affirmative, the relevant retry counter $\operatorname{RC}_m[\overline{k}]$ is checked against a suitable \emph{duplicate deferral threshold} ($D_{th}$).
If $\operatorname{RC}_m[\overline{k}]\ge D_{th}$ (that is, $\operatorname{F}_h$ has already experienced at least $D_{th}$ transmission attempts on channel $\overline{k}$) a duplicate of $\operatorname{F}_h$ is also sent by $\rsta{}_m[k]$.
Otherwise, the next frame $\operatorname{F}_{h^+}$, following $\operatorname{F}_h$ in $\operatorname{Q}_m^\mathit{tx}[k]$, is extracted from that queue and fed into the MAC of $\rsta{}_m[k]$.
Conceptually, this corresponds to swapping frames $\operatorname{F}_h$ and $\operatorname{F}_{h^+}$ at the head of $\operatorname{Q}_m^\mathit{tx}[k]$.
As shown in \cite{2014-ETFA-DDD}, setting $D_{th}=0$ disables DDD and reverts the system to pure RDA.

DDD adoption increases the RSTA throughput, since redundant channels can be exploited to send different packets at the same time.
As long as most transmissions are successful (this happens when channels suffer from mild interference and disturbance), the overall throughput is almost doubled.
Conversely, when the quality of channels worsens and a higher number of retransmissions is experienced, which exceeds $D_{th}$, reliability is preserved by means of frequency/time diversity.
Of course, leaving a channel idle is pointless when pending frames have real-time constraints, so deferral should not take place if $\operatorname{F}_h$ is the only frame in $\operatorname{Q}_m^\mathit{tx}[k]$.
In this case, the best option is sending a copy of $\operatorname{F}_h$ also on BSS[$k$].

The extension of DDD to triplex or higher redundancy ($R_{\operatorname{N}}\geq 3)$ is not difficult.
A solution is requiring that queued frames be suitably flagged.
In practice, each frame is set as \emph{ready} at its insertion in the transmission queue (unless $D_{th}=0$, which causes the frames to be marked \emph{undeferrable}).
When the frame is moved from the queue to a MAC, its state is switched to \emph{deferrable}.
As soon as $\operatorname{RC}$ in the MAC of any sub-STA $\rsta{}_m[g]$ reaches the threshold $D_{th}$ (that is, when $\operatorname{RC}_m[g]\geq D_{th}$), the frame being sent by that MAC is flagged as \emph{undeferrable}.
If the RSTA adopts separate transmission buffers, state changes have to be performed for all the frame copies stored in the different queues.

Whenever the MAC of any sub-STA in $\rsta{}_m$ becomes idle, the first \emph{undeferrable} frame is selected for transmission.
If there are no undeferrable frames, the first \emph{ready} frame is taken. 
Otherwise, when all the queued frames are marked \emph{deferrable}, the first one is transmitted anyway to reduce latency.
Whatever the case, frames that were previously discarded by any MAC of $\rsta{}_m$ shall not be sent again on the same BSS.
This has to be checked explicitly only if the RSTA relies on a single transmission buffer $\operatorname{Q}_m^\mathit{tx}$.
Conversely, frames discarded by $\rsta{}_m[k]$ are removed automatically from $\operatorname{Q}_m^\mathit{tx}[k]$ by the relevant MAC.
A brief synoptic of the operations carried out by the different DA approaches is reported in Table~\ref{tab:TbD}.

\subsection{Quality of Service}

Improving communication by means of seamless redundancy at the Wi-Fi physical layer  certainly helps
with ensuring true real-time behavior, but typically it is not sufficient on its own.
Mechanisms have to be included in the MAC layer as well, so that STAs (or RSTAs) can coordinate their access to the shared medium, hence lowering interference among devices belonging to the same system as much as possible.
To this extent, the point coordination function (PCF) was introduced in the very first IEEE 802.11 specification.
Later, it was noticeably improved with the hybrid coordination function (HCF) and, particularly, the HCF controlled channel access (HCCA).
A modified version of PCF is currently adopted by IWLAN whereas, to the best of our knowledge, HCCA is not supported by commercial devices yet.
In both cases a coordinator takes care of explicitly querying STAs, so as to ensure controlled and collision-free network access.

While advisable, the joint use of \wred{} and deterministic MAC mechanisms is not completely trivial.
In fact, cross-layer optimizations (which complement those carried out by the RDA and PDA mechanisms) should be devised in order to maximize benefits.
For this reason, in this paper we focus on a much simpler way to achieve QoS in Wi-Fi, that is, the enhanced distributed channel access (EDCA).
EDCA was originally conceived to support multimedia traffic.
Several solutions have been proposed in the recent literature, which foresee the use of the access categories (AC) defined by EDCA to improve communication determinism in soft real-time wireless control systems as well \cite{2007-TII-vas,2011-LCN-vas,2010-TII-802.11e}.

QoS can be easily added to \wred{} if EDCA is used in place of DCF.
This means that, in each sub-STA, traffic is split into four ACs, namely voice (AC\_VO), video (AC\_VI), best effort (AC\_BE), and background (AC\_BK), in decreasing priority order.
Each AC is provided with its own transmission queue, and a suitable AC is selected for each QoS-frame, mainly depending on its timing constraints.
For instance, the voice category is often envisaged to convey process data for which the transmission latency should be kept as short as possible, 
whereas best-effort traffic could be employed for tasks with no particular requirements about time.

While the adoption of EDCA does not affect RDA, some changes may be needed to PDA. For instance, the $D_{th}$ threshold in DDD can be selected on a per-AC basis.
In this case, its values are denoted DTH[AC]. From a practical point of view DTH[AC\_VO] can be set to a low value (e.g., $0$ or $1$), to cope with time-critical data.
With this arrangement, parallel transmissions of multiple copies of the same frame are started as soon as possible on each channel.
By contrast, a higher value (e.g., $3$ or more) should be selected for DTH[AC\_BK].
In this way, different frames will be sent (as long as possible) on  different channels at the same time, thus maximizing the throughput for background traffic at the expenses of increased latency for those frames which experience retransmissions.
So that, for any given AC, every RSTA in the RBSS receives the same QoS, DTH[AC] values must be selected on a network-wide basis. 
For instance, they can be included in beacons sent by the AP, in the same way as the other AC parameters, e.g., the arbitration interframe space number (AIFSN[AC]).

Besides DDD, other PDA schemes can be devised, which rely on more complex heuristics based on, e.g., queuing times. 
However, this requires  different sets of parameters with respect to DTH[AC].
The definition of additional PDA mechanisms, as well as the negotiation of their parameters in the network, are very complex matters and deserve further investigation.

\subsection{Frame reordering}
In transmission schemes that exploit diversity (and particularly PDA), frames may be received by the target node(s) in a different order than they were generated on the sender.
This behavior is not the same as conventional Wi-Fi, and may introduce problems in upper-layer protocols.
LRE can deal with this issue on the receiver side, and reorder frames before they are relayed to the upper layers through the logical link control (LLC) interface.
A single \emph{receive buffer} ($\operatorname{Q}_m^\mathit{rx}$) is then shared among all the sub-STAs of  $\rsta{}_m$.
Moreover, in order to grant compatibility among different \wred{} implementations, the receiver must not rely on the presence of any DA mechanism on the transmitter side.

The reordering procedure works as follow.
Received frames are inserted in $\operatorname{Q}_m^\mathit{rx}$ in their unique identifier order (that is the order followed by the DL-user on the transmitter side).
Frames received out of order introduce gaps in the buffer, and if a frame copy is already present in $\operatorname{Q}_m^\mathit{rx}$ a duplicate is not created.
$\operatorname{Q}_m^\mathit{rx}$ operates according to the sliding window principle: let $\operatorname{F}_x$ be the last frame correctly delivered to the DL-user on $\rsta{}_m$ and 
$\operatorname{F}_{x+1}$ the frame that was queued on the transmitter side of the redundant link (by the DL-user on the source RSTA) immediately after $\operatorname{F}_{x}$.
When $\operatorname{F}_{x+1}$ becomes available in  $\operatorname{Q}_m^\mathit{rx}$, it is extracted, together with all the \emph{subsequent} frames in $\operatorname{Q}_m^\mathit{rx}$ with contiguous identifiers (i.e., $\operatorname{F}_{x+2}$, $\operatorname{F}_{x+3}$, etc.), and orderly delivered to the local DL-user.
This corresponds to moving the sliding window forward up to the next gap (or until the queue is empty).

If a frame is dropped on all channels (quite unlikely event) a permanent gap appears, which would prevent the sliding window from advancing, thus blocking frame transfers. 
Timeouts can be used to circumvent this problem. 
In principle, a different timeout is started every time the \emph{first} copy of a frame is received and stored in $\operatorname{Q}_m^\mathit{rx}$.
If the timeout for any frame $\operatorname{F}_{y}$ expires, that frame ($\operatorname{F}_{y}$), all its \emph{preceding} frames (i.e., frames $\operatorname{F}_{z}$, $z < y$, which were queued for transmission before $\operatorname{F}_{y}$), and all the frames following $\operatorname{F}_{y}$ with contiguous identifiers (i.e., $\operatorname{F}_{y+1}$, $\operatorname{F}_{y+2}$, etc.) are extracted from $\operatorname{Q}_m^\mathit{rx}$ and delivered to the DL-user on $\rsta{}_m$.
Frames preceding $\operatorname{F}_{y}$ which have not been received yet are definitively lost.

Many aspects of reordering resemble PRP---and, to some extent, the mechanism to deal with block acknowledgment in Wi-Fi---but implementation and optimization details are not discussed here.
In practice, three basic policies can be adopted for the delivery of frames to the DL-user, that is:
\begin{enumerate}
\item 
  \emph{Ordered}: frames are reordered as explained above, and then delivered orderly to the upper layers. In this case, latency may increase appreciably.
\item 
  \emph{Unordered}: frames are delivered at once after duplicates have been removed. Implementation is very simple, but its behavior differs from conventional Wi-Fi.
\item 
  \emph{Not-unordered}: out-of-order late frames are discarded (this option is suitable for producer-consumer schemes only). Latency decreases, but the amount of dropped frames grows larger. Optimization is possible also on the sender side: for example, as soon as a new frame $\operatorname{F}_i$ is queued, all pending transmissions for earlier frames $\operatorname{F}_j$, $j<i$, can be aborted. As this policy is rather peculiar, it will not be considered in the following.
\end{enumerate}

\subsection{Security and fault tolerance}
Seamless redundancy protocols like PRP and \wred{} may suffer from issues related to security aspects.
For instance, an attacker may inject frames with wrong sequence numbers and, by exploiting the mechanism that discards duplicates in receivers, force them to throw away all the correct copies of the related packets.
The net effect resembles a denial of service (DoS) attack.
However, very effective security protocols exist for Wi-Fi, like the Wi-Fi protected access (WPA and WPA2), which are currently included in every commercial device.
They use a message integrity check (MIC), which prevents the receiving RSTAs from accepting fake frames forged by attackers.
If secure wireless connections are set up in \wred{}, LREs in receivers are certain to operate on authentic messages.

Another aspect that is very important in several application fields---including industrial plants---is fault tolerance \cite{2015-TII-FaultTolerance}.
Ensuring fault-tolerant communications among devices in distributed control systems can be achieved by resorting to, e.g., PRP.
This requires that two distinct, parallel networks are deployed.
If the system comprises Wi-Fi links, the related wireless devices have to be replicated as well, and should operate on different channels as in PoW.
A point that needs to be stressed about \wred{} is that, unlike PRP, it is not meant to enhance system reliability against \emph{permanent} faults in devices and cables, but to improve communication quality over IEEE 802.11 wireless links---in terms of packet losses and transmission latencies---in the presence of \emph{temporary} channel disturbance and interference.
\wred{} is noticeably cheaper than PRP, as it does not require full network duplication.
All it is needed is upgrading Wi-Fi equipment to \wred{}.

\section{Performance Evaluation}\label{sec:PerfEval}
The performance of \wred{} has been first evaluated through network simulation and compared to non-redundant Wi-Fi (i.e., plain DCF).
Then, a comparison was made among DA mechanisms.
In the following, duplex redundancy is considered since it is the most cost-effective solution for real systems.

\subsection{Communication quality}
Two important indexes for communication quality in industrial scenarios are the \emph{packet loss ratio} (fraction of packets that never reach their destination) and the \emph{transmission latency} (time elapsed between the instant when a packet is queued by the transmitting user and the instant it becomes available to the receiving user).
Both indexes are measured at the application level: in the following, we will call ``\emph{packet}'' the piece of information to be exchanged (as seen above the redundancy layer), while ``\emph{frame}'' refers to what is actually sent on air.

Hard real-time applications require that delays experienced by packets never exceed predefined deadlines.
Softer requirements, typical of non-time-critical systems, may refer to  statistical properties of latency in \emph{steady-state} conditions.
In this case, system behavior is correct as long as constraints are not violated for statistical indexes.
For instance, an upper bound can be defined for the \emph{fraction} of packets that are allowed to exceed a given deadline.
Equivalently, a specific  latency \emph{percentile} can be requested to be lower than such a deadline.

Typical statistical indexes for the transmission latency $d$ include its \emph{average} value ($\overline{d}$), \emph{standard deviation} ($\sigma_d$), \emph{minimum} ($d_{min}$) and \emph{maximum} ($d_{max}$) values, as well as percentiles.
In all experiments $d_{min}$ was equal to the theoretical transmission time over the air ($\unit[38]{\mu s}$).
Conversely, $d_{max}$ values were found to be little significant for our purposes, as they could vary appreciably also in similar experiments.
For this reason we took into account the 95, 99, and 99.9 percentiles ($d_{p95}$, $d_{p99}$, and $d_{p99.9}$, respectively) in place of $d_{max}$. Obviously, $d_{p95} \leq d_{p99} \leq d_{p99.9} \leq d_{max}$.

Latency $d_i$ experienced by any given packet can be expressed as the sum of 3 components: the \emph{queuing time} ($d_Q$) spent in $\operatorname{Q}^\mathit{tx}$ before the frame transmission begins, the \emph{transmission time} ($d_T$), which includes possible retransmissions, until the frame is delivered successfully to its destination, and the \emph{reordering time} ($d_R$) possibly spent in $\operatorname{Q}^\mathit{rx}$ before the packet is relayed to the upper layers.
The same contributions can be found in the corresponding average values.

With respect to packet delivery, the packet loss ratio $P_{lost}$ takes into account packets that have been definitely lost (i.e., packets whose copies were eventually dropped on both channels because of transmission buffer overruns, timeouts in the reordering buffer, or the retry limit being exceeded).
$P_{d=d_{min}}$ is the  fraction of packets incurring in delays due only to propagation on air and processing by adapters, whereas $P_{d>d_{min}}=1-P_{d=d_{min}}$ concerns packets for which $d>d_{min}$ (i.e., packets which have been either lost or further delayed because of buffer queuing, retransmissions, or deferral by the MAC).
Similarly, $P_{d>1}$, $P_{d>10}$, and $P_{d>100}$ represent \emph{deadline miss ratios}, and refer to packets whose delivery took more than $1$, $10$, and $\unit[100]{ms}$, respectively.
Obviously $P_{d>d_{min}}\geq P_{d>1}\geq P_{d>10}\geq P_{d>100}\geq P_{lost}$.

\subsection{Testbed structure}
\begin{figure}
  \centering
  \includegraphics[width=\columnwidth]{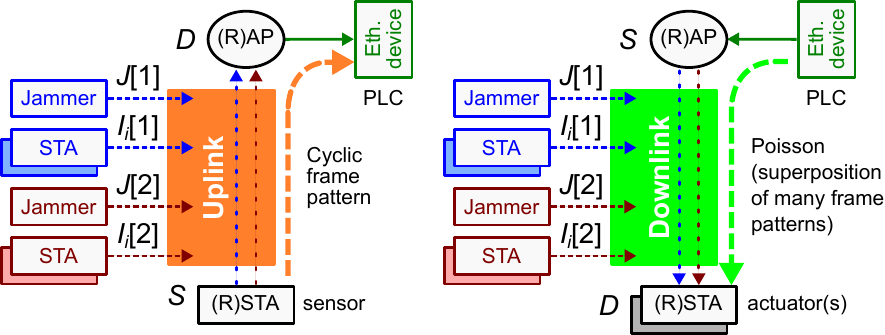}
  \caption{Structure of the testbed used for simulations (uplink and downlink).}
  \label{fig:testbed}
\end{figure}
Simulation experiments were carried out by means of a popular network modeler, and the testbed sketched in Fig.~\ref{fig:testbed} was used to evaluate the communication quality on a single (simplex or duplex) unidirectional wireless \emph{link}.
The experimental setup consists of:
\begin{itemize}
\item 
  one \emph{source} node (\emph{S}), which generates unicast packets to be sent on the link according to predefined rules;
\item 
  one \emph{destination} node (\emph{D}), which receives packets and computes statistical indexes of interest;
\item 
  a number of \emph{interfering} nodes (\emph{I}$_i$), which produce concurrent bulk traffic on wireless channels;
\item 
  two \emph{jammer} nodes (\emph{J}), which inject noise pulses on channels according to the Gilbert-Elliot model.
\end{itemize}
In case of simplex DCF links, \emph{S} and \emph{D} (as well as interferers) were based on the conventional IEEE 802.11 simulator models.
The same is true for \bwred{} (practically similar to \pow{}), whereas RSTA models were purposely developed for \emph{S} and \emph{D} to deal with DA mechanisms in \wred{}.

DCF enables a set of STAs in an IBSS to share the network bandwidth fairly, but the most common setup in real applications are by far infrastructure networks. 
APs act as \emph{hubs}, which collect and relay all frames among nodes (both wireless and wired).
Thus, the amount of traffic sent by the AP is noticeably higher than for ordinary STAs.
When the BSS experiences temporary congestion, the AP may easily become the network bottleneck.
This is a well-known problem \cite{2010-TII-802.11e}, and many improvements to basic DCF (i.e., transmission opportunities and frame aggregation \cite{2012-std-80211}) were primarily conceived bearing APs in mind.
From this point of view \wred{} is no exception: the RAP is the node where  frames may likely suffer from longer latencies and even get lost because of buffer overflows.

To cope with the RAP \emph{asymmetric} behavior, two different situations were considered in the experiments, dealing with transmissions from an RSTA to the RAP (\emph{uplink}) and viceversa (\emph{downlink}).

\subsubsection{Uplink transmissions}
in this case experiments somehow resemble  those presented in \cite{2014-ETFA-DDD}, with \emph{S} and \emph{D} acting as a generic RSTA and the RAP, respectively.
In a typical scenario, \emph{S} models a \emph{sensor} that sends its measured data to a \emph{controller} (e.g., a PLC connected to a wired segment) through the RAP.
The packet generation law is predictable and was selected to be  cyclical.
Since modern APs are frequently equipped with Gigabit Ethernet interfaces, outgoing frames from the RAP can be safely assumed not to be queued.
Moreover, transmission times over wires can also be neglected.
In theory, if the packet \emph{arrival rate} $\lambda$ on the \emph{S}$\rightarrow$\emph{D} link is kept below the available network capacity, cyclic generation does not cause frame queuing.
Practically, interfering traffic and disturbance have to reach quite high levels to affect the link communication quality in a tangible way.
As shown in the following, transmission latency is due, for a non-negligible part, to the frame reordering procedure carried out by the RAP.

\subsubsection{Downlink transmissions}
In this network setup \emph{S} acts as the RAP whereas \emph{D} models \emph{multiple} RSTAs (e.g., a set of remote \emph{actuators}).
It is worth noting that this is possible since all downlinks  originating from \emph{S} share the same DCF protocol automaton, and each link is granted the same average communication quality.
The packet generation law for \emph{S} consists of the superposition of several non-synchronized cyclical streams with different transmission periods.
To produce general results as much as possible, this was approximated by a Poisson process with arrival rate $\lambda$, so that the whole \emph{S}$\rightarrow$\emph{D} traffic can be characterized by means of a single parameter.

Differently from the uplink experiments, frame reordering was not taken into account. Actually, reordering is carried out separately for each downlink by the relevant target RSTA.
If the number of RSTAs modeled by \emph{D} is large enough, the order of the frames sent to any specific device is unlikely affected by DA mechanisms in a significant way. In practice, the contribution of reordering to latency is much smaller than in the uplink case, and hence  it can be safely neglected.
\begin{table}[b]
  \caption{Simulation parameters for the source traffic (\emph{S}$\rightarrow$\emph{D} link).}
  \label{tab:SD_param}
  \begin{center}
    \scriptsize
    \begin{tabular}{l||ccc}
      Parameter                                         & \emph{c}(1)    & \emph{e}(1)     & \emph{e}(0.5) \\
      \hline \hline
      Direction                                         & \emph{uplink}  & \emph{downlink} & \emph{downlink} \\
      Packet payload size                               & $\unit[50]{B}$ & $\unit[50]{B}$  & $\unit[50]{B}$ \\
      Packet generation law 				& cyclic         & exponential     & exponential \\
      Mean generation time ($\overline{T}=1/\lambda{}$)	& $\unit[1]{ms}$ & $\unit[1]{ms}$  & $\unit[0.5]{ms}$\\
      \hline
    \end{tabular}
  \end{center}
\end{table}
\begin{table}[b]
  \caption{Simulation parameters for the surrounding environment.}
  \label{tab:ENV_param}
  \begin{center}
    \scriptsize
    \begin{tabular}{c|l||cc}
               & Parameter                                    & \emph{Benign} ($E_B$) & \emph{Hostile} ($E_H$) \\
      \hline \hline
               & Frames per burst                             & $700$ & $700$ \\
               & Frame payload size                           & $\unit[1500]{B}$ & $\unit[1500]{B}$ \\
      Interf.  & Frame generation period 	              & $\unit[500]{\mu s}$ & $\unit[500]{\mu s}$ \\
               & Gap between bursts                           & $\operatorname{Exp}(\unit[1]{s})$ & $\operatorname{Exp}(\unit[1]{s})$ \\
               & Number of interferers                        & $2$ & $4$ \\
               & Overall interfering traffic ($\Lambda$)      & $\geq36\%$ & $\geq72\%$ \\
      \hline
               & Step duration                                & $\unit[1]{\mu s}$  & $\unit[1]{\mu s}$ \\
               & Trans. prob. good$\rightarrow$bad ($p_{gb})$ & $1.74\cdot10^{-4}$ & $1.74\cdot10^{-4}$ \\
      Disturb. & Trans. prob. bad$\rightarrow$good ($p_{bg}$) & $1.74\cdot10^{-2}$ & $1.74\cdot10^{-3}$ \\
               & Bit error prob. good state ($p_g$)           & $0$                & $0$ \\
               & Bit error prob. bad state ($p_b$)            & $7.5\cdot10^{-2}$  & $7.5\cdot10^{-2}$ \\
      \hline
    \end{tabular}
  \end{center}
\end{table}
\begin{table*}
  \caption{Results of the steady-state analysis with disturbance and interference (Wi-Fi, \bwred, and RDA mechanisms).}
  \label{tab:res_steady}
  \scriptsize
  \begin{center}
    \tabcolsep=0.15cm
    \begin{tabular}{cc|c||l||cc|cccc|ccccc|cc}
      \multicolumn{3}{c||}{\textbf{Environment}} & \textbf{Redund.} & \multicolumn{6}{c|}{\textbf{Statistics about packet transmission latency [$\unit[]{ms}$]}} & \multicolumn{5}{c|}{\textbf{Percentages about transmission latencies and failures [$\unit[]{\tcperthousand}$]}} & \multicolumn{2}{c}{\textbf{Mean queue size}}\\
      \multicolumn{3}{c||}{\textbf{and \emph{S}$\rightarrow$\emph{D} traffic}} &  \textbf{approach} & $\overline{d} $ & $\sigma_{d}$ & $d_{p95}$ & $d_{p99}$ & $d_{p99.9}$ & $d_{max}$ & $P_{d>d_{min}}$ & $P_{d>1}$ & $P_{d>10}$ & $P_{d>100}$ & $P_{lost}$ & $\overline{q}[1]$ & $\overline{q}[2]$ \\
      \hline \hline
      \multirow{11}{*}{\begin{sideways}Benign environment $E_B$\end{sideways}}
      & \multirow{4}{*}{\begin{sideways}\emph{Uplink}\end{sideways}}
      & \multirow{4}{*}{\begin{sideways}\emph{c}~(1)\end{sideways}}
        & \std{}   & 4.34 & 13.78   &  31.9 & 72.3 &  116.6 & 218    &   475 & 263 & 93.6 & 2.38    & 0.0001  &   4.62 &   -    \\
      & & & \pow{}   & 0.482 & 2.93    &  1.391 & 8.50 & 45.9 & 102.6       &   226 & 69.6 & 9.03 & 0.0013 & 0       &   4.67 & 4.60   \\
      & & & RDA/Q  & 0.141 & 0.394    &  0.649 & 1.575 & 4.69 & 31.9          &   202 & 24.1 & 0.1477 & 0.0  & 0       &   0.53 & 0.53   \\
      & & & RDA/R  & 0.113 & 0.237    &  0.524 & 1.114 & 2.41 & 20.3          &   194 & 13.22 & 0.0045 & 0.0 & 0       &   0.48 & 0.48   \\
      \cline{2-17}
      & \multirow{8}{*}{\begin{sideways}\emph{Downlink}\end{sideways}}
      & \multirow{4}{*}{\begin{sideways}\emph{e}~(1)\end{sideways}}
        & \std{}   &  5.16 & 14.89  &  34.0 & 78.8 & 128.6 & 271.3   &   636 & 355 & 105.4 & 3.64     & 0.0002 &   5.72 &   -    \\
      & & & \pow{}   &  0.806 & 3.43   &  3.41 & 12.14 & 51.1 & 113.9     &   451 & 136.4 & 12.49 & 0.0092 & 0      &   5.84 & 5.75   \\
      & & & RDA/Q  &  0.261 & 0.658   &  1.119 & 2.85 & 7.45 & 39.9         &   421 & 58.2 & 0.473 & 0.0     & 0      &   0.79 & 0.79   \\
      & & & RDA/R  &  0.212 & 0.413   &  0.909 & 1.986 & 4.13 & 21.3         &   416 & 42.6 & 0.025 & 0.0     & 0      &   0.72 & 0.72   \\
      \cline{3-17}
      & & \multirow{4}{*}{\begin{sideways}\emph{e}~(0.5)\end{sideways}}

      & \std{}   & 105.3 & 115.6  &  342 & 466 & 589 & 750  &  894 & 717 & 670 & 408    & 21.73    & 197.0 &     -   \\ 
      & & & \pow{}   & 46.6 & 69.2  &  192.2 & 293 & 427 & 713  &  812 & 543 & 473 & 183.2  & 0.622     & 194.9 & 196.4  \\
      & & & RDA/Q  &  2.84 & 9.33   &   16.41 & 49.2 & 98.4  & 199.0  &  658 & 215 & 75.4 & 0.921 & 0          & 6.11 & 6.08      \\
      & & & RDA/R  &  1.107 & 3.72   &    5.08 & 17.99 & 46.7 & 99.2      &  647 & 167.4 & 23.4 & 0.0 & 0          & 2.72 & 2.71      \\
      \hline \hline
      \multirow{11}{*}{\begin{sideways}Hostile environment $E_H$\end{sideways}}
      & \multirow{4}{*}{\begin{sideways}\emph{Uplink}\end{sideways}}
      & \multirow{4}{*}{\begin{sideways}\emph{c}~(1)\end{sideways}}
        & \std{}   & 54.7 & 88.3   & 237 & 396 &  624 & 894  &   788 & 664 & 478 & 205        & 0.1803       &   54.71 &   -  \\
      & & & \pow{}   & 15.53 & 36.6   &  93.4 & 173.0 & 295 & 587  &   626 & 446 & 232 & 44.0       & 0            & 55.65 & 55.52   \\
      & & & RDA/Q  & 1.126 & 4.39     &   4.31 & 20.0 & 58.6 & 145.8     &   520 & 177.3 & 23.2 & 0.1020  & 0            & 1.38 & 1.38     \\
      & & & RDA/R  & 0.435 & 1.091     &   1.721 & 4.54 & 13.37 & 53.2        &   500 & 109.8 & 2.06 & 0.0     & 0           & 0.69 & 0.69      \\
      \cline{2-17}
      & \multirow{8}{*}{\begin{sideways}\emph{Downlink}\end{sideways}}
      & \multirow{4}{*}{\begin{sideways}\emph{e}~(1)\end{sideways}}
        & \std{}   &  57.5 & 89.8  & 241 & 404 & 613 & 1203  &   856 & 723 & 509 & 216     &  0.208    & 57.97 &   -    \\
      & & & \pow{}   &  16.96 & 37.9  &  97.2 & 180.2 & 306 & 644   &   738 & 523 & 258 & 47.4    & 0          & 58.21 & 58.35  \\
      & & & RDA/Q  &   1.642 & 5.27   &  6.85 & 25.7 & 64.9 & 166.0       &   656 & 269 & 34.0 & 0.1893 & 0          & 2.15 & 2.14    \\
      & & & RDA/R  &   0.759 & 1.658   &  3.07 & 7.47 & 18.97 & 82.2         &   644 & 210 & 5.32 & 0.0    & 0          & 1.25 & 1.25    \\
      \cline{3-17}
      & & \multirow{4}{*}{\begin{sideways}\emph{e}~(0.5)\end{sideways}}
      & \std{}   & 338 & 204 &  697 & 834 & 976 & 1242 &   999 & 735 & 724 & 662   & 245  & 469.6 &   -       \\
      & & & \pow{}   & 263 & 180.7 &  600 & 752 & 911 & 1320 &   998 & 893 & 871 & 753   & 67.2   & 467.8 & 467.7    \\
      & & & RDA/Q  &  48.9 & 71.7  &  201 & 312 & 429 & 713  &   987 & 671 & 529 & 184.9 & 1.845    & 95.61 & 95.48      \\
      & & & RDA/R  &  21.9 & 39.5  &  107.1 & 182 & 276 & 407  &   985 & 585 & 369 & 58.1  & 0.0421    & 43.53 & 43.46      \\
      \hline
    \end{tabular}
  \end{center}
\end{table*}

\subsection{Testbed configuration}
In each experiment \emph{S} was configured to send a large number of packets addressed to \emph{D}, with a payload size equal to $\unit[50]{B}$.
The period between subsequent packets was set to the constant value $\unit[1]{ms}$ in uplink simulations, whereas packet inter-arrival times were exponentially distributed for downlink experiments and two sub-cases were considered, with expected value $\overline{T}=1/\lambda{}$ equal to $\unit[0.5]{ms}$ and $\unit[1]{ms}$, respectively.
Overall, three kinds of \emph{S}$\rightarrow$\emph{D} traffic were taken into account, by varying the generation law---either \emph{cyclic} (\emph{c}) or \emph{exponential} (\emph{e})---and mean intertime ($\overline{T}$), as reported in Table~\ref{tab:SD_param}.
In the following they are denoted \emph{c}(1), \emph{e}(1), and \emph{e}(0.5).
Instead, two different conditions were considered for  interference and disturbance of the surrounding environment, that is \emph{benign} ($E_B$) and \emph{hostile} ($E_H$).
Parameters of interest are listed in Table~\ref{tab:ENV_param}.

Interfering traffic consists of a number of STAs that repeatedly generate bursts of $700$ frames.
For each burst, frames with a $\unit[1500]{B}$ payload are sent every $\unit[500]{\mu s}$.
Unlike \cite{2014-WFCS-WiRed} and \cite{2014-ETFA-DDD}, acknowledged transmissions were used, as they more faithfully model concurrent TCP/IP traffic over the air.
The inter-burst gap is an exponentially distributed random variable with mean value $\unit[1]{s}$.
Traffic produced on BSS[$k$] by each interfering node $I_i[k]$ (excluding retransmissions) is equal to $700\cdot(T_\mathit{TX}+T_\mathit{SIFS}+T_\mathit{ACK}+T_\mathit{DIFS})/(700\cdot\unit[500]{\mu s}+\unit[1]{s})\simeq\unit[18]{\%}$ of the available channel bandwidth, where $T_\mathit{TX}$ and $T_\mathit{ACK}$ are the transmission times of interfering and ACK frames ($\unit[254]{\mu s}$ and $\unit[34]{\mu s}$, respectively), while $T_\mathit{SIFS}$ and $T_\mathit{DIFS}$ are the short and distributed interframe space durations ($\unit[10]{\mu s}$ and $\unit[50]{\mu s}$, respectively, as nodes in the testbed were set for simplicity as part of an IBSS).
We used two interfering nodes for each channel in $E_B$ and four nodes in $E_H$.
In this way, the overall interfering traffic $\Lambda$ on each BSS was $\geq\unit[36]{\%}$ and $\geq\unit[72]{\%}$, respectively.

Disturbance was modeled independently on each channel according to the Gilbert-Elliott  model \cite{1960-gilbert, 1963-elliott}.
Indeed, other models exist \cite{2015-TWC-Tanghe}, which  describe the wireless spectrum more accurately when the environment is known to some extent.
However, the main purpose of this simulation campaign was comparing approaches based on seamless redundancy (including RDA and PDA techniques) with plain Wi-Fi in the same operating conditions.
A preliminary measurement campaign was run on a prototype PoW setup made up of real devices, deployed in both a lab environment (high interfering traffic) and an industrial plant (with high-current welding guns).
Results showed that the advantages of \wred{} over Wi-Fi mainly depend on whether channels are reasonably independent (which is the case of most real-world systems), while the specific nature of the surrounding environment plays a marginal role.
For these reasons, and in order to obtain generally applicable results, we opted for the simpler Gilbert-Elliott disturbance model.

As in \cite{2001-report-gilbertElliott}, 5 parameters are needed to this purpose, that is the transition probability from good to bad state ($p_{gb}$) and viceversa ($p_{bg}$), the bit error probability for good ($p_g$) and  bad state ($p_b$), and the step duration ($\unit[1]{\mu s}$).
Gilbert-Elliott parameters for $E_B$ are similar to those reported in \cite{2001-report-gilbertElliott} and resemble a \emph{typical} real-world Wi-Fi channel ($p_{gb}=1.74\cdot10^{-4}$, $p_{bg}=1.74\cdot10^{-2}$, $p_g=0$, and $p_b=7.5\cdot10^{-2}$).
To stress the proposed DA techniques, disturbance in $E_H$ was set about one order of magnitude higher than $E_B$ ($p_{bg}$ was decreased to $1.74\cdot10^{-3}$).
In the simulator, disturbance was achieved using one jammer per channel ($J[1]$ and $J[2]$).

Much more cases were actually analyzed in our simulation campaign, besides those described above, by considering alternate sets of parameters.
Only the most significant ones are presented in the following for space reasons.

\section{Results}\label{sec:Results}
RDA mechanisms were evaluated independently and compared to both plain DCF and \bwred{} (\pow{}). 
Then, a specific PDA technique (namely DDD) was considered and  the impact of the deferral threshold value on latencies analyzed.
\begin{figure*}
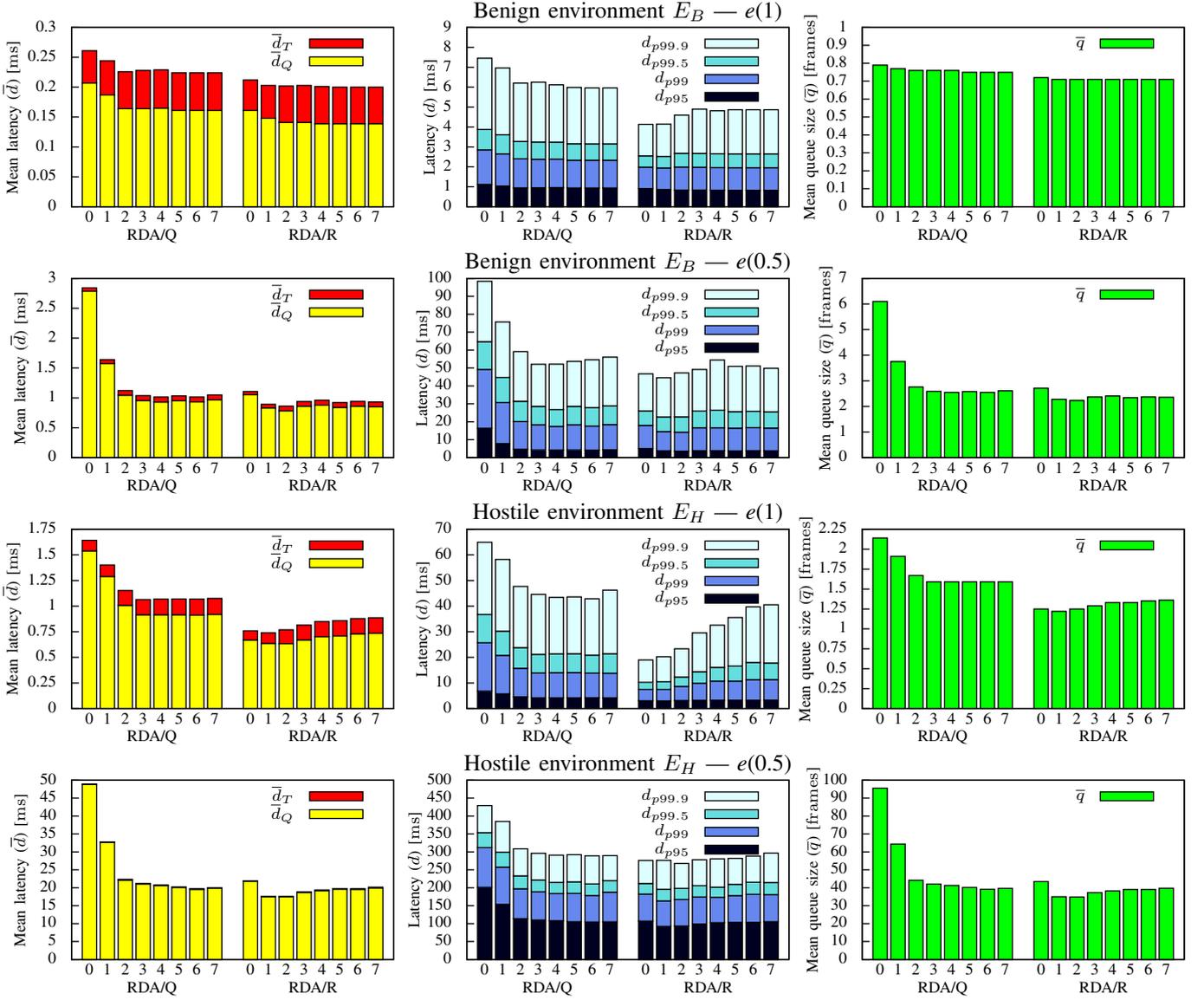

  \scriptsize
  \centering
  \include{FIG4-15-1127}
  \caption{Results for \wred{} with PDA (DDD): \emph{downlink} case (\emph{Poisson} \emph{S}$\rightarrow$\emph{D} traffic), RDA/R and RDA/Q (with and without ADCF).}
  \label{fig:exp_1}
\end{figure*}

\subsection{RDA mechanisms}
Table \ref{tab:res_steady} summarizes the statistical performance indexes for legacy (non-redundant) DCF, \pow{} (which, over a single redundant link, behaves the same as  \bwred{}, without any DA), and \wred{} (with either RDA/Q or RDA/R).

\subsubsection{Uplink transmissions}
In this case, the \emph{S}$\rightarrow$\emph{D} generation law is  \emph{c}(1).
In benign environmental conditions ($E_B$) \bwred{} provided a tenfold reduction of the mean transmission latency $\overline{d}$ over DCF, and the adoption of RDA improved its behavior further.
In hostile conditions ($E_H$) \bwred{} was no longer able to grant such a huge improvement, whereas RDA boosted performance noticeably again. 
The reason is that, in hostile environments RDA prevents queues from growing out of control.
The analysis of higher percentiles ($d_{p99.9}$) shows that  RDA always outperformed both DCF and \bwred{}.
This means that very few packets were excessively delayed  in the experiments.

By comparing RDA/Q and RDA/R, it is clear that the ability of the latter to abort transmissions at the MAC level brings significant advantages with respect to the simple removal of frames in $\operatorname{Q}^\mathit{tx}$ carried out by the former, in particular when the environment is hostile. 
This is due to the decrease in the overall network load when useless retransmissions are aborted.

\begin{figure*}
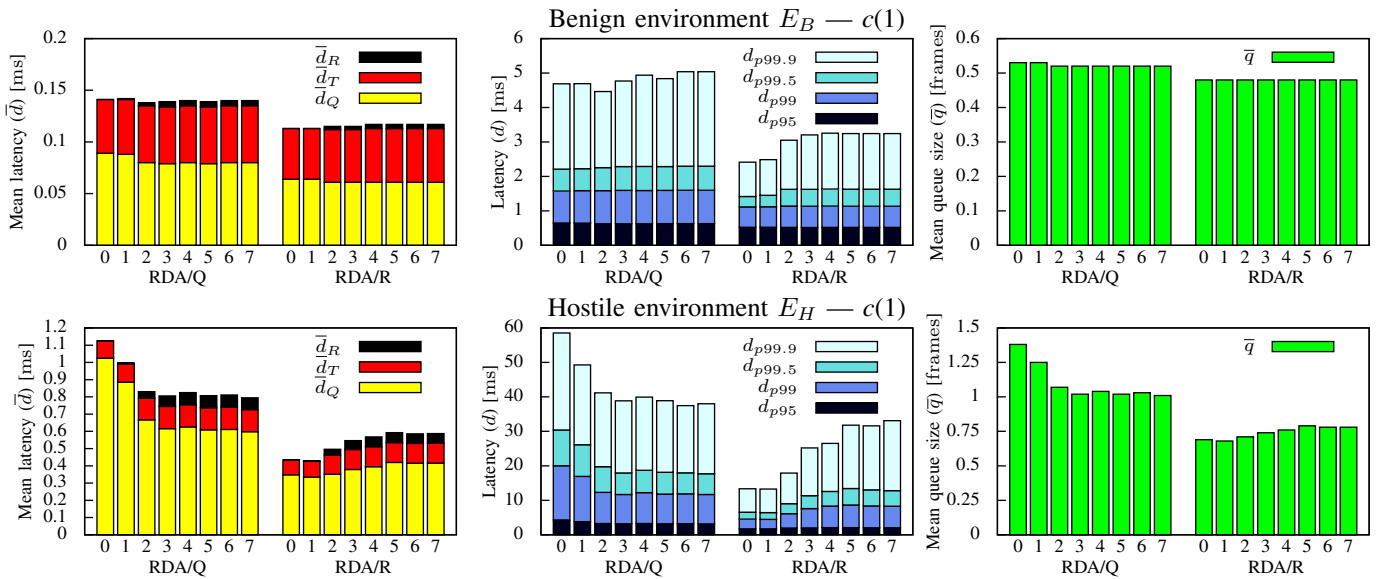

  \scriptsize
  \centering
  \include{FIG5-15-1127}
  \caption{Results for \wred{} with PDA (DDD): \emph{uplink} case (\emph{cyclic} \emph{S}$\rightarrow$\emph{D} traffic), RDA/R and RDA/Q (with and without ADCF).} 
  \label{fig:exp_2}
\end{figure*}

\subsubsection{Downlink transmissions}
Two different values were used for the \emph{S}$\rightarrow$\emph{D} generation rate $\lambda$, so that in \emph{e}(0.5) it was twice than in \emph{e}(1).
Since RDA is aimed at preventing duplicate transmissions, the available network bandwidth virtually doubles with respect to \bwred{}.
A comparison of the  mean  transmission latency $\overline{d}$ for \pow{} in \emph{e}(1) and RDA/R in \emph{e}(0.5) (i.e., $\unit[0.806]{ms}$ vs. $\unit[1.107]{ms}$ in $E_B$ and $\unit[16.96]{ms}$ vs. $\unit[21.9]{ms}$ in $E_H$) shows that RDA was able to improve the link throughput  by a factor close to two  with respect to \bwred{}.

Poisson arrivals with rate $\lambda$ close to the link throughput, as in \emph{e}(0.5), may cause $\operatorname{Q}^\mathit{tx}$ to fill up, and this behavior  is even more evident in hostile environment ($E_H$).
Consequently, both latency and the packet loss ratio increase.
Experiments confirmed that RDA, and in particular RDA/R, can provide substantial benefits to the fraction of dropped packets.
For instance, in the worst conditions---i.e., \emph{e}(0.5) and $E_H$---more than $24\%$ of packets were lost in conventional Wi-Fi (DCF);  basic redundant schemes like \pow{} reduced this fraction to less than $7\%$, while the adoption of DA mechanisms shrinked it down to $0.18\%$ (RDA/Q) and $0.004\%$ (RDA/R).
\begin{table}[b]
  \caption{Packet loss ratio, unordered (U) and ordered (O) policies.}
  \label{tab:exp_config}
  \begin{center}
    \footnotesize
    \begin{tabular}{c||cc|cc|cc|cc}
        & \multicolumn{8}{c}{Packet Loss Ratio ($P_{lost}$)\ \ \ [$\cdot 10^{-6}$]} \\
      \cline{2-9}
			& \multicolumn{4}{c|}{\textbf{$E_B$}} & \multicolumn{4}{c}{\textbf{$E_H$}} \\
      \cline{2-9}
       & \multicolumn{2}{c|}{RDA/Q} & \multicolumn{2}{c|}{RDA/R} & \multicolumn{2}{c|}{RDA/Q} & \multicolumn{2}{c}{RDA/R} \\
        PDA/$D_{th}$                & U & O & U & O & U & O & U & O \\
      \hline \hline
      DDD/0   & 0.0 & 0.0  & 0.0 & 0.0 & 0.0 & 0.0   & 0.0 & 0.0   \\
      DDD/1   & 0.0 & 0.1  & 0.0 & 0.0 & 0.0 & 1.3   & 0.0 & 1.5   \\
      DDD/2   & 0.0 & 0.1  & 0.0 & 0.3 & 0.0 & 14.4  & 0.0 & 13.8  \\
      DDD/3   & 0.0 & 1.4  & 0.0 & 0.7 & 0.0 & 131.5 & 0.0 & 118.9 \\
      DDD/4   & 0.0 & 7.2  & 0.0 & 4.9 & 0.0 & 331.7 & 0.0 & 269.2 \\
      DDD/5   & 0.0 & 8.5  & 0.0 & 5.7 & 0.0 & 345.2 & 0.0 & 287.8 \\
      DDD/6   & 0.0 & 10.3 & 0.0 & 5.7 & 0.0 & 352.3 & 0.0 & 280.8 \\
      DDD/7   & 0.0 & 10.3 & 0.0 & 5.7 & 0.0 & 335.8 & 0.0 & 269.1 \\
      \hline
    \end{tabular}
  \end{center}
\end{table}

\subsection{PDA mechanisms}
Figures \ref{fig:exp_1} and \ref{fig:exp_2} show the effect of adopting PDA  along with RDA in  \emph{downlink} and \emph{uplink} transmissions, respectively.
DDD was explicitly considered and the threshold value $D_{th}$  varied in the range from $0$ (DDD  disabled) to $7$ (a duplicate is sent only after the first copy of a frame has been discarded by the MAC on the other channel).

Three groups of diagrams are shown for each experiment.
The leftmost vertical group in each figure  concerns the mean latency $\overline{d}$: each bar highlights the different  contributions due to queuing ($\overline{d}_Q$), transmission ($\overline{d}_T$) and, for \emph{c}(1) only, reordering ($\overline{d}_R$).
The second group concerns percentiles, shown in terms of difference with respect to the predecessor.
Finally, the rightmost group depicts the mean queue size $\overline{q}$ (in redundant schemes, $\overline{q}=(\overline{q}[1]+\overline{q}[2])/2$).
Two sets of bars are included in each diagram, which concern RDA/Q (on the left) and RDA/R (on the right), respectively.

\subsubsection{Downlink transmissions}\label{sec:Downlink}
As can be seen in Fig.~\ref{fig:exp_1}, the mean latency $\overline{d}$ mainly depended on the time $d_Q$ spent by frames in the transmission buffer $\operatorname{Q}^\mathit{tx}$.
As a consequence, any DA mechanism that helps reducing the number of queued frames improves the overall link quality.
In all diagrams in Fig.~\ref{fig:exp_1}, RDA/R always performs better than RDA/Q.
The cost to be paid is that a modified MAC is needed, able to support some kind of ADCF.

The RDA/Q behavior can be effectively improved by selecting higher values for $D_{th}$, in particular when $\lambda$ approaches the mean channel throughput.
In fact, increasing $D_{th}$ reduces the amount of duplicates sent on air.
An optimal $D_{th}$ value in this case is around 3.
The adoption of DDD with RDA/R introduced slight improvements when $\lambda$ was high, as in \emph{e}(0.5), but only when $D_{th}=1$.
This is because, with duplex redundancy, such a choice enables scattering frame transmissions on the two channels.
When $D_{th}>1$ the retry counter and backoff duration are allowed to grow, and this worsens latency.

\subsubsection{Uplink transmissions}\label{sec:Uplink}
Similar considerations as above apply to results shown in Fig.~\ref{fig:exp_2}, where the number of queued packets is lower on average because packet generation is cyclic.
Consequently, benefits introduced by DDD are not as significant as in the case of Poisson arrivals.
As can be seen in the leftmost graphs, latency is due in part to reordering in the reception buffer,
especially in hostile environment ($E_H$), where the order of the received frames may be altered tangibly because of redundant transmission.

\subsubsection{Packet loss rate}\label{sec:packetloss}
Timeliness is useless if too many packets are dropped.
Table \ref{tab:exp_config} reports $P_{lost}$ values for \emph{c}(1) experiments (uplink).
The first row of the table shows that no packets were lost when RDA mechanisms were used alone ($D_{th}=0$ means that DDD was disabled).
The same held when the \emph{unordered} delivery policy was adopted.
Conversely, packets were dropped for higher $D_{th}$ values in the case of \emph{ordered} delivery. 
This occurs because of timeouts in the reordering block, which prevent packets from being stuck in $\operatorname{Q}^\mathit{rx}$ for more than $\unit[10]{ms}$.

\section{Conclusions}\label{sec:Conclusions}
A fundamental requirement in industrial applications is determinism.
Consequently, underlying communication systems must grant that all packets are delivered timely and correctly.
Less demanding real-time control systems also exist, where the above property has to hold for the vast majority of packets.
Unfortunately, wireless networks---and in particular Wi-Fi, which is considered the wireless counterpart of Ethernet---are not reliable enough, not even for the latter kind of applications.

As the wireless transmission medium is  not under complete control of system designers and managers, interference and disturbance can not be avoided.
For this reason, providing some ``quasi-deterministic'' overlay---implemented, e.g., as a software protocol layer on top of Wi-Fi---is often necessary, but usually not sufficient.
Seamless redundancy permits to face peculiarities of radio transmission and improves the wireless channel reliability.
Besides plain use of PRP and redundant Wi-Fi equipment, the adoption of duplicate avoidance mechanisms, which prevent the bandwidth from being wasted, improves the system behavior further.

In this paper, a unified framework has been provided to describe and categorize the most recent proposals that apply seamless link-level redundancy to Wi-Fi.
A thorough comparison of the available solutions has been carried out through simulation.
Results are  encouraging, and show that communication quality can be largely improved---up to orders of magnitude---by introducing small changes in  existing wireless devices (no alteration is needed in the physical layer and only minor modifications are requested to the MAC).
It is worth remarking that, at present, a large number of Wi-Fi devices are dual band, and many APs support simultaneous dual-band operations as well.
This means, that most hardware resources needed to support seamless redundancy are already there.

While mainly targeted at industrial, building, and home automation systems, \wred{} is very appealing also in other application scenarios, such as high-fidelity multimedia streaming.
For this reason, we believe that it should be incorporated in future versions of IEEE 802.11 standards.

\bibliographystyle{IEEEtran}
\bibliography{TII-15-1127}

\begin{IEEEbiography}%
[{\includegraphics[width=1in,height=1.25in,clip,keepaspectratio]
{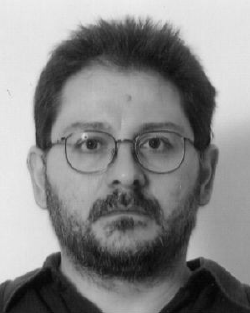}}]{Gianluca Cena} (SM'09) received the Laurea degree in electronic engineering and the Ph.D. degree in information and system engineering from the Politecnico di Torino, Turin, Italy, in 1991 and 1996,
respectively.

In 1995, he became an Assistant Professor in the Department of Computer Engineering, Politecnico di Torino.  Since 2005 he has been a Director of Research with the National Research Council of Italy (CNR), and in particular with the Institute of Electronics, Computer, and Telecommunication Engineering (IEIIT), Turin. His research interests include industrial communication systems and real-time networks. In these areas, he has coauthored more than 100 technical papers.

Dr. Cena served as Program Co-Chairman for the 2006 and 2008 editions of the IEEE Workshop on Factory Communication Systems, and has been an Associate Editor of the \textsc{IEEE Transactions on Industrial Informatics} since 2009.
\end{IEEEbiography}

\begin{IEEEbiography}%
[{\includegraphics[width=1in,height=1.25in,clip,keepaspectratio]
{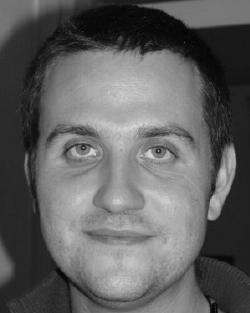}}]{Stefano Scanzio} (S'06-M'12) received the Laurea and Ph.D. degrees in computer science from Politecnico di Torino, Torino, Italy, in 2004 and 2008, respectively.

He was with the Department of Computer Engineering, Politecnico di Torino, from 2004 to 2009, where he was involved in research on speech recognition and, in particular, he has been active in classification methods and algorithms. Since 2009, he has been with the National Research Council of Italy (CNR), where he is a tenured technical Researcher with the Institute of Electronics, Computer and Telecommunication Engineering (IEIIT), Turin.

Dr. Scanzio teaches several courses on computer science at Politecnico di Torino. He has authored and co-authored several papers in international journals and conferences in the area of industrial communication systems and real-time networks.
\end{IEEEbiography}

\begin{IEEEbiography}%
[{\includegraphics[width=1in,height=1.25in,clip,keepaspectratio]
{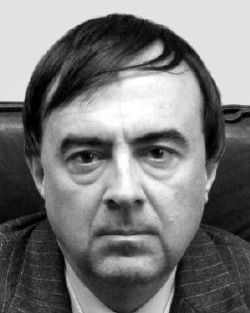}}]{Adriano Valenzano} (SM'09) received the Laurea degree in electronic engineering from Politecnico di Torino, Torino, Italy, in 1980. He is Director of Research with the National Research Council of Italy (CNR). He is currently with the Institute of Electronics, Computer and Telecommunication Engineering (IEIIT), Torino, Italy, where he is responsible for research concerning distributed computer systems, local area networks, and communication protocols. He has coauthored approximately 200 refereed journal and conference papers in the area of computer engineering.

Dr. Valenzano is the recipient of the 2013 IEEE IES and ABB Lifetime Contribution to Factory Automation Award. He also received, as a coauthor, the Best Paper Award presented at the Fifth and Eighth IEEE Workshops on Factory Communication Systems (WFCS 2004 and WFCS 2010). 

He has served as a technical referee for several international journals and conferences, also taking part in the program committees of international events of primary importance. Since 2007, he has been serving as an Associate Editor for the \textsc{IEEE Transactions on Industrial Informatics}.
\end{IEEEbiography}

\vfill

\end{document}

%% file: FIG4-15-1127.tex
% GNUPLOT: LaTeX picture with Postscript
\begingroup
  \makeatletter
  \providecommand\color[2][]{%
    \GenericError{(gnuplot) \space\space\space\@spaces}{%
      Package color not loaded in conjunction with
      terminal option `colourtext'%
    }{See the gnuplot documentation for explanation.%
    }{Either use 'blacktext' in gnuplot or load the package
      color.sty in LaTeX.}%
    \renewcommand\color[2][]{}%
  }%
  \providecommand\includegraphics[2][]{%
    \GenericError{(gnuplot) \space\space\space\@spaces}{%
      Package graphicx or graphics not loaded%
    }{See the gnuplot documentation for explanation.%
    }{The gnuplot epslatex terminal needs graphicx.sty or graphics.sty.}%
    \renewcommand\includegraphics[2][]{}%
  }%
  \providecommand\rotatebox[2]{#2}%
  \@ifundefined{ifGPcolor}{%
    \newif\ifGPcolor
    \GPcolorfalse
  }{}%
  \@ifundefined{ifGPblacktext}{%
    \newif\ifGPblacktext
    \GPblacktexttrue
  }{}%
  % define a \g@addto@macro without @ in the name:
  \let\gplgaddtomacro\g@addto@macro
  % define empty templates for all commands taking text:
  \gdef\gplbacktext{}%
  \gdef\gplfronttext{}%
  \makeatother
  \ifGPblacktext
    % no textcolor at all
    \def\colorrgb#1{}%
    \def\colorgray#1{}%
  \else
    % gray or color?
    \ifGPcolor
      \def\colorrgb#1{\color[rgb]{#1}}%
      \def\colorgray#1{\color[gray]{#1}}%
      \expandafter\def\csname LTw\endcsname{\color{white}}%
      \expandafter\def\csname LTb\endcsname{\color{black}}%
      \expandafter\def\csname LTa\endcsname{\color{black}}%
      \expandafter\def\csname LT0\endcsname{\color[rgb]{1,0,0}}%
      \expandafter\def\csname LT1\endcsname{\color[rgb]{0,1,0}}%
      \expandafter\def\csname LT2\endcsname{\color[rgb]{0,0,1}}%
      \expandafter\def\csname LT3\endcsname{\color[rgb]{1,0,1}}%
      \expandafter\def\csname LT4\endcsname{\color[rgb]{0,1,1}}%
      \expandafter\def\csname LT5\endcsname{\color[rgb]{1,1,0}}%
      \expandafter\def\csname LT6\endcsname{\color[rgb]{0,0,0}}%
      \expandafter\def\csname LT7\endcsname{\color[rgb]{1,0.3,0}}%
      \expandafter\def\csname LT8\endcsname{\color[rgb]{0.5,0.5,0.5}}%
    \else
      % gray
      \def\colorrgb#1{\color{black}}%
      \def\colorgray#1{\color[gray]{#1}}%
      \expandafter\def\csname LTw\endcsname{\color{white}}%
      \expandafter\def\csname LTb\endcsname{\color{black}}%
      \expandafter\def\csname LTa\endcsname{\color{black}}%
      \expandafter\def\csname LT0\endcsname{\color{black}}%
      \expandafter\def\csname LT1\endcsname{\color{black}}%
      \expandafter\def\csname LT2\endcsname{\color{black}}%
      \expandafter\def\csname LT3\endcsname{\color{black}}%
      \expandafter\def\csname LT4\endcsname{\color{black}}%
      \expandafter\def\csname LT5\endcsname{\color{black}}%
      \expandafter\def\csname LT6\endcsname{\color{black}}%
      \expandafter\def\csname LT7\endcsname{\color{black}}%
      \expandafter\def\csname LT8\endcsname{\color{black}}%
    \fi
  \fi
  \setlength{\unitlength}{0.0500bp}%
  \begin{picture}(10318.00,9070.00)%
    \gplgaddtomacro\gplbacktext{%
      \csname LTb\endcsname%
      \put(396,6970){\makebox(0,0)[r]{\strut{} 0}}%
      \put(396,7229){\makebox(0,0)[r]{\strut{} 0.05}}%
      \put(396,7488){\makebox(0,0)[r]{\strut{} 0.1}}%
      \put(396,7748){\makebox(0,0)[r]{\strut{} 0.15}}%
      \put(396,8007){\makebox(0,0)[r]{\strut{} 0.2}}%
      \put(396,8266){\makebox(0,0)[r]{\strut{} 0.25}}%
      \put(396,8525){\makebox(0,0)[r]{\strut{} 0.3}}%
      \put(682,6871){\makebox(0,0){\strut{}0}}%
      \put(836,6871){\makebox(0,0){\strut{}1}}%
      \put(990,6871){\makebox(0,0){\strut{}2}}%
      \put(1144,6871){\makebox(0,0){\strut{}3}}%
      \put(1298,6871){\makebox(0,0){\strut{}4}}%
      \put(1452,6871){\makebox(0,0){\strut{}5}}%
      \put(1606,6871){\makebox(0,0){\strut{}6}}%
      \put(1760,6871){\makebox(0,0){\strut{}7}}%
      \put(2069,6871){\makebox(0,0){\strut{}0}}%
      \put(2223,6871){\makebox(0,0){\strut{}1}}%
      \put(2377,6871){\makebox(0,0){\strut{}2}}%
      \put(2531,6871){\makebox(0,0){\strut{}3}}%
      \put(2685,6871){\makebox(0,0){\strut{}4}}%
      \put(2839,6871){\makebox(0,0){\strut{}5}}%
      \put(2993,6871){\makebox(0,0){\strut{}6}}%
      \put(3147,6871){\makebox(0,0){\strut{}7}}%
      \put(22,7747){\rotatebox{-270}{\makebox(0,0){\strut{}Mean latency ($\overline{d}$) [$\unit[]{ms}$]}}}%
      \put(1914,6552){\makebox(0,0){\strut{} }}%
    }%
    \gplgaddtomacro\gplfronttext{%
      \csname LTb\endcsname%
      \put(2446,8231){\makebox(0,0)[r]{\strut{}$\overline{d}_Q$}}%
      \csname LTb\endcsname%
      \put(2446,8385){\makebox(0,0)[r]{\strut{}$\overline{d}_T$}}%
      \csname LTb\endcsname%
      \put(2608,6717){\makebox(0,0){\strut{}RDA/R}}%
      \put(1221,6717){\makebox(0,0){\strut{}RDA/Q}}%
    }%
    \gplgaddtomacro\gplbacktext{%
      \csname LTb\endcsname%
      \put(396,4793){\makebox(0,0)[r]{\strut{} 0}}%
      \put(396,5052){\makebox(0,0)[r]{\strut{} 0.5}}%
      \put(396,5311){\makebox(0,0)[r]{\strut{} 1}}%
      \put(396,5571){\makebox(0,0)[r]{\strut{} 1.5}}%
      \put(396,5830){\makebox(0,0)[r]{\strut{} 2}}%
      \put(396,6089){\makebox(0,0)[r]{\strut{} 2.5}}%
      \put(396,6348){\makebox(0,0)[r]{\strut{} 3}}%
      \put(682,4694){\makebox(0,0){\strut{}0}}%
      \put(836,4694){\makebox(0,0){\strut{}1}}%
      \put(990,4694){\makebox(0,0){\strut{}2}}%
      \put(1144,4694){\makebox(0,0){\strut{}3}}%
      \put(1298,4694){\makebox(0,0){\strut{}4}}%
      \put(1452,4694){\makebox(0,0){\strut{}5}}%
      \put(1606,4694){\makebox(0,0){\strut{}6}}%
      \put(1760,4694){\makebox(0,0){\strut{}7}}%
      \put(2069,4694){\makebox(0,0){\strut{}0}}%
      \put(2223,4694){\makebox(0,0){\strut{}1}}%
      \put(2377,4694){\makebox(0,0){\strut{}2}}%
      \put(2531,4694){\makebox(0,0){\strut{}3}}%
      \put(2685,4694){\makebox(0,0){\strut{}4}}%
      \put(2839,4694){\makebox(0,0){\strut{}5}}%
      \put(2993,4694){\makebox(0,0){\strut{}6}}%
      \put(3147,4694){\makebox(0,0){\strut{}7}}%
      \put(88,5570){\rotatebox{-270}{\makebox(0,0){\strut{}Mean latency ($\overline{d}$) [$\unit[]{ms}$]}}}%
      \put(1914,4375){\makebox(0,0){\strut{} }}%
    }%
    \gplgaddtomacro\gplfronttext{%
      \csname LTb\endcsname%
      \put(2446,6054){\makebox(0,0)[r]{\strut{}$\overline{d}_Q$}}%
      \csname LTb\endcsname%
      \put(2446,6208){\makebox(0,0)[r]{\strut{}$\overline{d}_T$}}%
      \csname LTb\endcsname%
      \put(2608,4540){\makebox(0,0){\strut{}RDA/R}}%
      \put(1221,4540){\makebox(0,0){\strut{}RDA/Q}}%
    }%
    \gplgaddtomacro\gplbacktext{%
      \csname LTb\endcsname%
      \put(396,2616){\makebox(0,0)[r]{\strut{} 0}}%
      \put(396,2838){\makebox(0,0)[r]{\strut{} 0.25}}%
      \put(396,3061){\makebox(0,0)[r]{\strut{} 0.5}}%
      \put(396,3283){\makebox(0,0)[r]{\strut{} 0.75}}%
      \put(396,3505){\makebox(0,0)[r]{\strut{} 1}}%
      \put(396,3727){\makebox(0,0)[r]{\strut{} 1.25}}%
      \put(396,3950){\makebox(0,0)[r]{\strut{} 1.5}}%
      \put(396,4172){\makebox(0,0)[r]{\strut{} 1.75}}%
      \put(682,2517){\makebox(0,0){\strut{}0}}%
      \put(836,2517){\makebox(0,0){\strut{}1}}%
      \put(990,2517){\makebox(0,0){\strut{}2}}%
      \put(1144,2517){\makebox(0,0){\strut{}3}}%
      \put(1298,2517){\makebox(0,0){\strut{}4}}%
      \put(1452,2517){\makebox(0,0){\strut{}5}}%
      \put(1606,2517){\makebox(0,0){\strut{}6}}%
      \put(1760,2517){\makebox(0,0){\strut{}7}}%
      \put(2069,2517){\makebox(0,0){\strut{}0}}%
      \put(2223,2517){\makebox(0,0){\strut{}1}}%
      \put(2377,2517){\makebox(0,0){\strut{}2}}%
      \put(2531,2517){\makebox(0,0){\strut{}3}}%
      \put(2685,2517){\makebox(0,0){\strut{}4}}%
      \put(2839,2517){\makebox(0,0){\strut{}5}}%
      \put(2993,2517){\makebox(0,0){\strut{}6}}%
      \put(3147,2517){\makebox(0,0){\strut{}7}}%
      \put(22,3394){\rotatebox{-270}{\makebox(0,0){\strut{}Mean latency ($\overline{d}$) [$\unit[]{ms}$]}}}%
      \put(1914,2198){\makebox(0,0){\strut{} }}%
    }%
    \gplgaddtomacro\gplfronttext{%
      \csname LTb\endcsname%
      \put(2446,3878){\makebox(0,0)[r]{\strut{}$\overline{d}_Q$}}%
      \csname LTb\endcsname%
      \put(2446,4032){\makebox(0,0)[r]{\strut{}$\overline{d}_T$}}%
      \csname LTb\endcsname%
      \put(2608,2363){\makebox(0,0){\strut{}RDA/R}}%
      \put(1221,2363){\makebox(0,0){\strut{}RDA/Q}}%
    }%
    \gplgaddtomacro\gplbacktext{%
      \csname LTb\endcsname%
      \put(396,440){\makebox(0,0)[r]{\strut{} 0}}%
      \put(396,596){\makebox(0,0)[r]{\strut{} 5}}%
      \put(396,751){\makebox(0,0)[r]{\strut{} 10}}%
      \put(396,907){\makebox(0,0)[r]{\strut{} 15}}%
      \put(396,1062){\makebox(0,0)[r]{\strut{} 20}}%
      \put(396,1218){\makebox(0,0)[r]{\strut{} 25}}%
      \put(396,1373){\makebox(0,0)[r]{\strut{} 30}}%
      \put(396,1529){\makebox(0,0)[r]{\strut{} 35}}%
      \put(396,1684){\makebox(0,0)[r]{\strut{} 40}}%
      \put(396,1840){\makebox(0,0)[r]{\strut{} 45}}%
      \put(396,1995){\makebox(0,0)[r]{\strut{} 50}}%
      \put(682,341){\makebox(0,0){\strut{}0}}%
      \put(836,341){\makebox(0,0){\strut{}1}}%
      \put(990,341){\makebox(0,0){\strut{}2}}%
      \put(1144,341){\makebox(0,0){\strut{}3}}%
      \put(1298,341){\makebox(0,0){\strut{}4}}%
      \put(1452,341){\makebox(0,0){\strut{}5}}%
      \put(1606,341){\makebox(0,0){\strut{}6}}%
      \put(1760,341){\makebox(0,0){\strut{}7}}%
      \put(2069,341){\makebox(0,0){\strut{}0}}%
      \put(2223,341){\makebox(0,0){\strut{}1}}%
      \put(2377,341){\makebox(0,0){\strut{}2}}%
      \put(2531,341){\makebox(0,0){\strut{}3}}%
      \put(2685,341){\makebox(0,0){\strut{}4}}%
      \put(2839,341){\makebox(0,0){\strut{}5}}%
      \put(2993,341){\makebox(0,0){\strut{}6}}%
      \put(3147,341){\makebox(0,0){\strut{}7}}%
      \put(88,1217){\rotatebox{-270}{\makebox(0,0){\strut{}Mean latency ($\overline{d}$) [$\unit[]{ms}$]}}}%
      \put(1914,22){\makebox(0,0){\strut{} }}%
    }%
    \gplgaddtomacro\gplfronttext{%
      \csname LTb\endcsname%
      \put(2446,1701){\makebox(0,0)[r]{\strut{}$\overline{d}_Q$}}%
      \csname LTb\endcsname%
      \put(2446,1855){\makebox(0,0)[r]{\strut{}$\overline{d}_T$}}%
      \csname LTb\endcsname%
      \put(2608,187){\makebox(0,0){\strut{}RDA/R}}%
      \put(1221,187){\makebox(0,0){\strut{}RDA/Q}}%
    }%
    \gplgaddtomacro\gplbacktext{%
      \csname LTb\endcsname%
      \put(3800,6970){\makebox(0,0)[r]{\strut{} 0}}%
      \put(3800,7143){\makebox(0,0)[r]{\strut{} 1}}%
      \put(3800,7316){\makebox(0,0)[r]{\strut{} 2}}%
      \put(3800,7488){\makebox(0,0)[r]{\strut{} 3}}%
      \put(3800,7661){\makebox(0,0)[r]{\strut{} 4}}%
      \put(3800,7834){\makebox(0,0)[r]{\strut{} 5}}%
      \put(3800,8007){\makebox(0,0)[r]{\strut{} 6}}%
      \put(3800,8179){\makebox(0,0)[r]{\strut{} 7}}%
      \put(3800,8352){\makebox(0,0)[r]{\strut{} 8}}%
      \put(3800,8525){\makebox(0,0)[r]{\strut{} 9}}%
      \put(4086,6871){\makebox(0,0){\strut{}0}}%
      \put(4240,6871){\makebox(0,0){\strut{}1}}%
      \put(4394,6871){\makebox(0,0){\strut{}2}}%
      \put(4548,6871){\makebox(0,0){\strut{}3}}%
      \put(4703,6871){\makebox(0,0){\strut{}4}}%
      \put(4857,6871){\makebox(0,0){\strut{}5}}%
      \put(5011,6871){\makebox(0,0){\strut{}6}}%
      \put(5165,6871){\makebox(0,0){\strut{}7}}%
      \put(5473,6871){\makebox(0,0){\strut{}0}}%
      \put(5627,6871){\makebox(0,0){\strut{}1}}%
      \put(5781,6871){\makebox(0,0){\strut{}2}}%
      \put(5935,6871){\makebox(0,0){\strut{}3}}%
      \put(6090,6871){\makebox(0,0){\strut{}4}}%
      \put(6244,6871){\makebox(0,0){\strut{}5}}%
      \put(6398,6871){\makebox(0,0){\strut{}6}}%
      \put(6552,6871){\makebox(0,0){\strut{}7}}%
      \put(3624,7747){\rotatebox{-270}{\makebox(0,0){\strut{}Latency ($d$) [$\unit[]{ms}$]}}}%
      \put(5319,6552){\makebox(0,0){\strut{} }}%
      \put(5319,8657){\makebox(0,0){\strut{}\normalsize Benign environment $E_B$ --- \emph{e}(1)}}%
    }%
    \gplgaddtomacro\gplfronttext{%
      \csname LTb\endcsname%
      \put(5851,7923){\makebox(0,0)[r]{\strut{}$d_{p95}$}}%
      \csname LTb\endcsname%
      \put(5851,8077){\makebox(0,0)[r]{\strut{}$d_{p99}$}}%
      \csname LTb\endcsname%
      \put(5851,8231){\makebox(0,0)[r]{\strut{}$d_{p99.5}$}}%
      \csname LTb\endcsname%
      \put(5851,8385){\makebox(0,0)[r]{\strut{}$d_{p99.9}$}}%
      \csname LTb\endcsname%
      \put(6013,6717){\makebox(0,0){\strut{}RDA/R}}%
      \put(4626,6717){\makebox(0,0){\strut{}RDA/Q}}%
    }%
    \gplgaddtomacro\gplbacktext{%
      \csname LTb\endcsname%
      \put(3800,4793){\makebox(0,0)[r]{\strut{} 0}}%
      \put(3800,4949){\makebox(0,0)[r]{\strut{} 10}}%
      \put(3800,5104){\makebox(0,0)[r]{\strut{} 20}}%
      \put(3800,5260){\makebox(0,0)[r]{\strut{} 30}}%
      \put(3800,5415){\makebox(0,0)[r]{\strut{} 40}}%
      \put(3800,5571){\makebox(0,0)[r]{\strut{} 50}}%
      \put(3800,5726){\makebox(0,0)[r]{\strut{} 60}}%
      \put(3800,5882){\makebox(0,0)[r]{\strut{} 70}}%
      \put(3800,6037){\makebox(0,0)[r]{\strut{} 80}}%
      \put(3800,6193){\makebox(0,0)[r]{\strut{} 90}}%
      \put(3800,6348){\makebox(0,0)[r]{\strut{} 100}}%
      \put(4086,4694){\makebox(0,0){\strut{}0}}%
      \put(4240,4694){\makebox(0,0){\strut{}1}}%
      \put(4394,4694){\makebox(0,0){\strut{}2}}%
      \put(4548,4694){\makebox(0,0){\strut{}3}}%
      \put(4703,4694){\makebox(0,0){\strut{}4}}%
      \put(4857,4694){\makebox(0,0){\strut{}5}}%
      \put(5011,4694){\makebox(0,0){\strut{}6}}%
      \put(5165,4694){\makebox(0,0){\strut{}7}}%
      \put(5473,4694){\makebox(0,0){\strut{}0}}%
      \put(5627,4694){\makebox(0,0){\strut{}1}}%
      \put(5781,4694){\makebox(0,0){\strut{}2}}%
      \put(5935,4694){\makebox(0,0){\strut{}3}}%
      \put(6090,4694){\makebox(0,0){\strut{}4}}%
      \put(6244,4694){\makebox(0,0){\strut{}5}}%
      \put(6398,4694){\makebox(0,0){\strut{}6}}%
      \put(6552,4694){\makebox(0,0){\strut{}7}}%
      \put(3558,5570){\rotatebox{-270}{\makebox(0,0){\strut{}Latency ($d$) [$\unit[]{ms}$]}}}%
      \put(5319,4375){\makebox(0,0){\strut{} }}%
      \put(5319,6480){\makebox(0,0){\strut{}\normalsize Benign environment $E_B$ --- \emph{e}(0.5)}}%
    }%
    \gplgaddtomacro\gplfronttext{%
      \csname LTb\endcsname%
      \put(5851,5746){\makebox(0,0)[r]{\strut{}$d_{p95}$}}%
      \csname LTb\endcsname%
      \put(5851,5900){\makebox(0,0)[r]{\strut{}$d_{p99}$}}%
      \csname LTb\endcsname%
      \put(5851,6054){\makebox(0,0)[r]{\strut{}$d_{p99.5}$}}%
      \csname LTb\endcsname%
      \put(5851,6208){\makebox(0,0)[r]{\strut{}$d_{p99.9}$}}%
      \csname LTb\endcsname%
      \put(6013,4540){\makebox(0,0){\strut{}RDA/R}}%
      \put(4626,4540){\makebox(0,0){\strut{}RDA/Q}}%
    }%
    \gplgaddtomacro\gplbacktext{%
      \csname LTb\endcsname%
      \put(3800,2616){\makebox(0,0)[r]{\strut{} 0}}%
      \put(3800,2838){\makebox(0,0)[r]{\strut{} 10}}%
      \put(3800,3061){\makebox(0,0)[r]{\strut{} 20}}%
      \put(3800,3283){\makebox(0,0)[r]{\strut{} 30}}%
      \put(3800,3505){\makebox(0,0)[r]{\strut{} 40}}%
      \put(3800,3727){\makebox(0,0)[r]{\strut{} 50}}%
      \put(3800,3950){\makebox(0,0)[r]{\strut{} 60}}%
      \put(3800,4172){\makebox(0,0)[r]{\strut{} 70}}%
      \put(4086,2517){\makebox(0,0){\strut{}0}}%
      \put(4240,2517){\makebox(0,0){\strut{}1}}%
      \put(4394,2517){\makebox(0,0){\strut{}2}}%
      \put(4548,2517){\makebox(0,0){\strut{}3}}%
      \put(4703,2517){\makebox(0,0){\strut{}4}}%
      \put(4857,2517){\makebox(0,0){\strut{}5}}%
      \put(5011,2517){\makebox(0,0){\strut{}6}}%
      \put(5165,2517){\makebox(0,0){\strut{}7}}%
      \put(5473,2517){\makebox(0,0){\strut{}0}}%
      \put(5627,2517){\makebox(0,0){\strut{}1}}%
      \put(5781,2517){\makebox(0,0){\strut{}2}}%
      \put(5935,2517){\makebox(0,0){\strut{}3}}%
      \put(6090,2517){\makebox(0,0){\strut{}4}}%
      \put(6244,2517){\makebox(0,0){\strut{}5}}%
      \put(6398,2517){\makebox(0,0){\strut{}6}}%
      \put(6552,2517){\makebox(0,0){\strut{}7}}%
      \put(3558,3394){\rotatebox{-270}{\makebox(0,0){\strut{}Latency ($d$) [$\unit[]{ms}$]}}}%
      \put(5319,2198){\makebox(0,0){\strut{} }}%
      \put(5319,4304){\makebox(0,0){\strut{}\normalsize Hostile environment $E_H$ --- \emph{e}(1)}}%
    }%
    \gplgaddtomacro\gplfronttext{%
      \csname LTb\endcsname%
      \put(5851,3570){\makebox(0,0)[r]{\strut{}$d_{p95}$}}%
      \csname LTb\endcsname%
      \put(5851,3724){\makebox(0,0)[r]{\strut{}$d_{p99}$}}%
      \csname LTb\endcsname%
      \put(5851,3878){\makebox(0,0)[r]{\strut{}$d_{p99.5}$}}%
      \csname LTb\endcsname%
      \put(5851,4032){\makebox(0,0)[r]{\strut{}$d_{p99.9}$}}%
      \csname LTb\endcsname%
      \put(6013,2363){\makebox(0,0){\strut{}RDA/R}}%
      \put(4626,2363){\makebox(0,0){\strut{}RDA/Q}}%
    }%
    \gplgaddtomacro\gplbacktext{%
      \csname LTb\endcsname%
      \put(3800,440){\makebox(0,0)[r]{\strut{} 0}}%
      \put(3800,596){\makebox(0,0)[r]{\strut{} 50}}%
      \put(3800,751){\makebox(0,0)[r]{\strut{} 100}}%
      \put(3800,907){\makebox(0,0)[r]{\strut{} 150}}%
      \put(3800,1062){\makebox(0,0)[r]{\strut{} 200}}%
      \put(3800,1218){\makebox(0,0)[r]{\strut{} 250}}%
      \put(3800,1373){\makebox(0,0)[r]{\strut{} 300}}%
      \put(3800,1529){\makebox(0,0)[r]{\strut{} 350}}%
      \put(3800,1684){\makebox(0,0)[r]{\strut{} 400}}%
      \put(3800,1840){\makebox(0,0)[r]{\strut{} 450}}%
      \put(3800,1995){\makebox(0,0)[r]{\strut{} 500}}%
      \put(4086,341){\makebox(0,0){\strut{}0}}%
      \put(4240,341){\makebox(0,0){\strut{}1}}%
      \put(4394,341){\makebox(0,0){\strut{}2}}%
      \put(4548,341){\makebox(0,0){\strut{}3}}%
      \put(4703,341){\makebox(0,0){\strut{}4}}%
      \put(4857,341){\makebox(0,0){\strut{}5}}%
      \put(5011,341){\makebox(0,0){\strut{}6}}%
      \put(5165,341){\makebox(0,0){\strut{}7}}%
      \put(5473,341){\makebox(0,0){\strut{}0}}%
      \put(5627,341){\makebox(0,0){\strut{}1}}%
      \put(5781,341){\makebox(0,0){\strut{}2}}%
      \put(5935,341){\makebox(0,0){\strut{}3}}%
      \put(6090,341){\makebox(0,0){\strut{}4}}%
      \put(6244,341){\makebox(0,0){\strut{}5}}%
      \put(6398,341){\makebox(0,0){\strut{}6}}%
      \put(6552,341){\makebox(0,0){\strut{}7}}%
      \put(3492,1217){\rotatebox{-270}{\makebox(0,0){\strut{}Latency ($d$) [$\unit[]{ms}$]}}}%
      \put(5319,22){\makebox(0,0){\strut{} }}%
      \put(5319,2127){\makebox(0,0){\strut{}\normalsize Hostile environment $E_H$ --- \emph{e}(0.5)}}%
    }%
    \gplgaddtomacro\gplfronttext{%
      \csname LTb\endcsname%
      \put(5851,1393){\makebox(0,0)[r]{\strut{}$d_{p95}$}}%
      \csname LTb\endcsname%
      \put(5851,1547){\makebox(0,0)[r]{\strut{}$d_{p99}$}}%
      \csname LTb\endcsname%
      \put(5851,1701){\makebox(0,0)[r]{\strut{}$d_{p99.5}$}}%
      \csname LTb\endcsname%
      \put(5851,1855){\makebox(0,0)[r]{\strut{}$d_{p99.9}$}}%
      \csname LTb\endcsname%
      \put(6013,187){\makebox(0,0){\strut{}RDA/R}}%
      \put(4626,187){\makebox(0,0){\strut{}RDA/Q}}%
    }%
    \gplgaddtomacro\gplbacktext{%
      \csname LTb\endcsname%
      \put(7205,6970){\makebox(0,0)[r]{\strut{} 0}}%
      \put(7205,7126){\makebox(0,0)[r]{\strut{} 0.1}}%
      \put(7205,7281){\makebox(0,0)[r]{\strut{} 0.2}}%
      \put(7205,7437){\makebox(0,0)[r]{\strut{} 0.3}}%
      \put(7205,7592){\makebox(0,0)[r]{\strut{} 0.4}}%
      \put(7205,7748){\makebox(0,0)[r]{\strut{} 0.5}}%
      \put(7205,7903){\makebox(0,0)[r]{\strut{} 0.6}}%
      \put(7205,8059){\makebox(0,0)[r]{\strut{} 0.7}}%
      \put(7205,8214){\makebox(0,0)[r]{\strut{} 0.8}}%
      \put(7205,8370){\makebox(0,0)[r]{\strut{} 0.9}}%
      \put(7205,8525){\makebox(0,0)[r]{\strut{} 1}}%
      \put(7491,6871){\makebox(0,0){\strut{}0}}%
      \put(7645,6871){\makebox(0,0){\strut{}1}}%
      \put(7799,6871){\makebox(0,0){\strut{}2}}%
      \put(7953,6871){\makebox(0,0){\strut{}3}}%
      \put(8107,6871){\makebox(0,0){\strut{}4}}%
      \put(8261,6871){\makebox(0,0){\strut{}5}}%
      \put(8415,6871){\makebox(0,0){\strut{}6}}%
      \put(8569,6871){\makebox(0,0){\strut{}7}}%
      \put(8878,6871){\makebox(0,0){\strut{}0}}%
      \put(9032,6871){\makebox(0,0){\strut{}1}}%
      \put(9186,6871){\makebox(0,0){\strut{}2}}%
      \put(9340,6871){\makebox(0,0){\strut{}3}}%
      \put(9494,6871){\makebox(0,0){\strut{}4}}%
      \put(9648,6871){\makebox(0,0){\strut{}5}}%
      \put(9802,6871){\makebox(0,0){\strut{}6}}%
      \put(9956,6871){\makebox(0,0){\strut{}7}}%
      \put(6897,7747){\rotatebox{-270}{\makebox(0,0){\strut{}Mean queue size ($\overline{q}$) [$\unit[]{frames}$]}}}%
      \put(8723,6552){\makebox(0,0){\strut{} }}%
    }%
    \gplgaddtomacro\gplfronttext{%
      \csname LTb\endcsname%
      \put(9255,8385){\makebox(0,0)[r]{\strut{}$\overline{q}$}}%
      \csname LTb\endcsname%
      \put(9417,6717){\makebox(0,0){\strut{}RDA/R}}%
      \put(8030,6717){\makebox(0,0){\strut{}RDA/Q}}%
    }%
    \gplgaddtomacro\gplbacktext{%
      \csname LTb\endcsname%
      \put(7205,4793){\makebox(0,0)[r]{\strut{} 0}}%
      \put(7205,5015){\makebox(0,0)[r]{\strut{} 1}}%
      \put(7205,5237){\makebox(0,0)[r]{\strut{} 2}}%
      \put(7205,5459){\makebox(0,0)[r]{\strut{} 3}}%
      \put(7205,5682){\makebox(0,0)[r]{\strut{} 4}}%
      \put(7205,5904){\makebox(0,0)[r]{\strut{} 5}}%
      \put(7205,6126){\makebox(0,0)[r]{\strut{} 6}}%
      \put(7205,6348){\makebox(0,0)[r]{\strut{} 7}}%
      \put(7491,4694){\makebox(0,0){\strut{}0}}%
      \put(7645,4694){\makebox(0,0){\strut{}1}}%
      \put(7799,4694){\makebox(0,0){\strut{}2}}%
      \put(7953,4694){\makebox(0,0){\strut{}3}}%
      \put(8107,4694){\makebox(0,0){\strut{}4}}%
      \put(8261,4694){\makebox(0,0){\strut{}5}}%
      \put(8415,4694){\makebox(0,0){\strut{}6}}%
      \put(8569,4694){\makebox(0,0){\strut{}7}}%
      \put(8878,4694){\makebox(0,0){\strut{}0}}%
      \put(9032,4694){\makebox(0,0){\strut{}1}}%
      \put(9186,4694){\makebox(0,0){\strut{}2}}%
      \put(9340,4694){\makebox(0,0){\strut{}3}}%
      \put(9494,4694){\makebox(0,0){\strut{}4}}%
      \put(9648,4694){\makebox(0,0){\strut{}5}}%
      \put(9802,4694){\makebox(0,0){\strut{}6}}%
      \put(9956,4694){\makebox(0,0){\strut{}7}}%
      \put(7029,5570){\rotatebox{-270}{\makebox(0,0){\strut{}Mean queue size ($\overline{q}$) [$\unit[]{frames}$]}}}%
      \put(8723,4375){\makebox(0,0){\strut{} }}%
    }%
    \gplgaddtomacro\gplfronttext{%
      \csname LTb\endcsname%
      \put(9255,6208){\makebox(0,0)[r]{\strut{}$\overline{q}$}}%
      \csname LTb\endcsname%
      \put(9417,4540){\makebox(0,0){\strut{}RDA/R}}%
      \put(8030,4540){\makebox(0,0){\strut{}RDA/Q}}%
    }%
    \gplgaddtomacro\gplbacktext{%
      \csname LTb\endcsname%
      \put(7205,2616){\makebox(0,0)[r]{\strut{} 0}}%
      \put(7205,2789){\makebox(0,0)[r]{\strut{} 0.25}}%
      \put(7205,2962){\makebox(0,0)[r]{\strut{} 0.5}}%
      \put(7205,3135){\makebox(0,0)[r]{\strut{} 0.75}}%
      \put(7205,3308){\makebox(0,0)[r]{\strut{} 1}}%
      \put(7205,3480){\makebox(0,0)[r]{\strut{} 1.25}}%
      \put(7205,3653){\makebox(0,0)[r]{\strut{} 1.5}}%
      \put(7205,3826){\makebox(0,0)[r]{\strut{} 1.75}}%
      \put(7205,3999){\makebox(0,0)[r]{\strut{} 2}}%
      \put(7205,4172){\makebox(0,0)[r]{\strut{} 2.25}}%
      \put(7491,2517){\makebox(0,0){\strut{}0}}%
      \put(7645,2517){\makebox(0,0){\strut{}1}}%
      \put(7799,2517){\makebox(0,0){\strut{}2}}%
      \put(7953,2517){\makebox(0,0){\strut{}3}}%
      \put(8107,2517){\makebox(0,0){\strut{}4}}%
      \put(8261,2517){\makebox(0,0){\strut{}5}}%
      \put(8415,2517){\makebox(0,0){\strut{}6}}%
      \put(8569,2517){\makebox(0,0){\strut{}7}}%
      \put(8878,2517){\makebox(0,0){\strut{}0}}%
      \put(9032,2517){\makebox(0,0){\strut{}1}}%
      \put(9186,2517){\makebox(0,0){\strut{}2}}%
      \put(9340,2517){\makebox(0,0){\strut{}3}}%
      \put(9494,2517){\makebox(0,0){\strut{}4}}%
      \put(9648,2517){\makebox(0,0){\strut{}5}}%
      \put(9802,2517){\makebox(0,0){\strut{}6}}%
      \put(9956,2517){\makebox(0,0){\strut{}7}}%
      \put(6857,3394){\rotatebox{-270}{\makebox(0,0){\strut{}Mean queue size ($\overline{q}$) [$\unit[]{frames}$]}}}%
      \put(8723,2198){\makebox(0,0){\strut{} }}%
    }%
    \gplgaddtomacro\gplfronttext{%
      \csname LTb\endcsname%
      \put(9255,4032){\makebox(0,0)[r]{\strut{}$\overline{q}$}}%
      \csname LTb\endcsname%
      \put(9417,2363){\makebox(0,0){\strut{}RDA/R}}%
      \put(8030,2363){\makebox(0,0){\strut{}RDA/Q}}%
    }%
    \gplgaddtomacro\gplbacktext{%
      \csname LTb\endcsname%
      \put(7205,440){\makebox(0,0)[r]{\strut{} 0}}%
      \put(7205,596){\makebox(0,0)[r]{\strut{} 10}}%
      \put(7205,751){\makebox(0,0)[r]{\strut{} 20}}%
      \put(7205,907){\makebox(0,0)[r]{\strut{} 30}}%
      \put(7205,1062){\makebox(0,0)[r]{\strut{} 40}}%
      \put(7205,1218){\makebox(0,0)[r]{\strut{} 50}}%
      \put(7205,1373){\makebox(0,0)[r]{\strut{} 60}}%
      \put(7205,1529){\makebox(0,0)[r]{\strut{} 70}}%
      \put(7205,1684){\makebox(0,0)[r]{\strut{} 80}}%
      \put(7205,1840){\makebox(0,0)[r]{\strut{} 90}}%
      \put(7205,1995){\makebox(0,0)[r]{\strut{} 100}}%
      \put(7491,341){\makebox(0,0){\strut{}0}}%
      \put(7645,341){\makebox(0,0){\strut{}1}}%
      \put(7799,341){\makebox(0,0){\strut{}2}}%
      \put(7953,341){\makebox(0,0){\strut{}3}}%
      \put(8107,341){\makebox(0,0){\strut{}4}}%
      \put(8261,341){\makebox(0,0){\strut{}5}}%
      \put(8415,341){\makebox(0,0){\strut{}6}}%
      \put(8569,341){\makebox(0,0){\strut{}7}}%
      \put(8878,341){\makebox(0,0){\strut{}0}}%
      \put(9032,341){\makebox(0,0){\strut{}1}}%
      \put(9186,341){\makebox(0,0){\strut{}2}}%
      \put(9340,341){\makebox(0,0){\strut{}3}}%
      \put(9494,341){\makebox(0,0){\strut{}4}}%
      \put(9648,341){\makebox(0,0){\strut{}5}}%
      \put(9802,341){\makebox(0,0){\strut{}6}}%
      \put(9956,341){\makebox(0,0){\strut{}7}}%
      \put(6923,1217){\rotatebox{-270}{\makebox(0,0){\strut{}Mean queue size ($\overline{q}$) [$\unit[]{frames}$]}}}%
      \put(8723,22){\makebox(0,0){\strut{} }}%
    }%
    \gplgaddtomacro\gplfronttext{%
      \csname LTb\endcsname%
      \put(9255,1855){\makebox(0,0)[r]{\strut{}$\overline{q}$}}%
      \csname LTb\endcsname%
      \put(9417,187){\makebox(0,0){\strut{}RDA/R}}%
      \put(8030,187){\makebox(0,0){\strut{}RDA/Q}}%
    }%
    \gplbacktext
    \put(0,0){\includegraphics{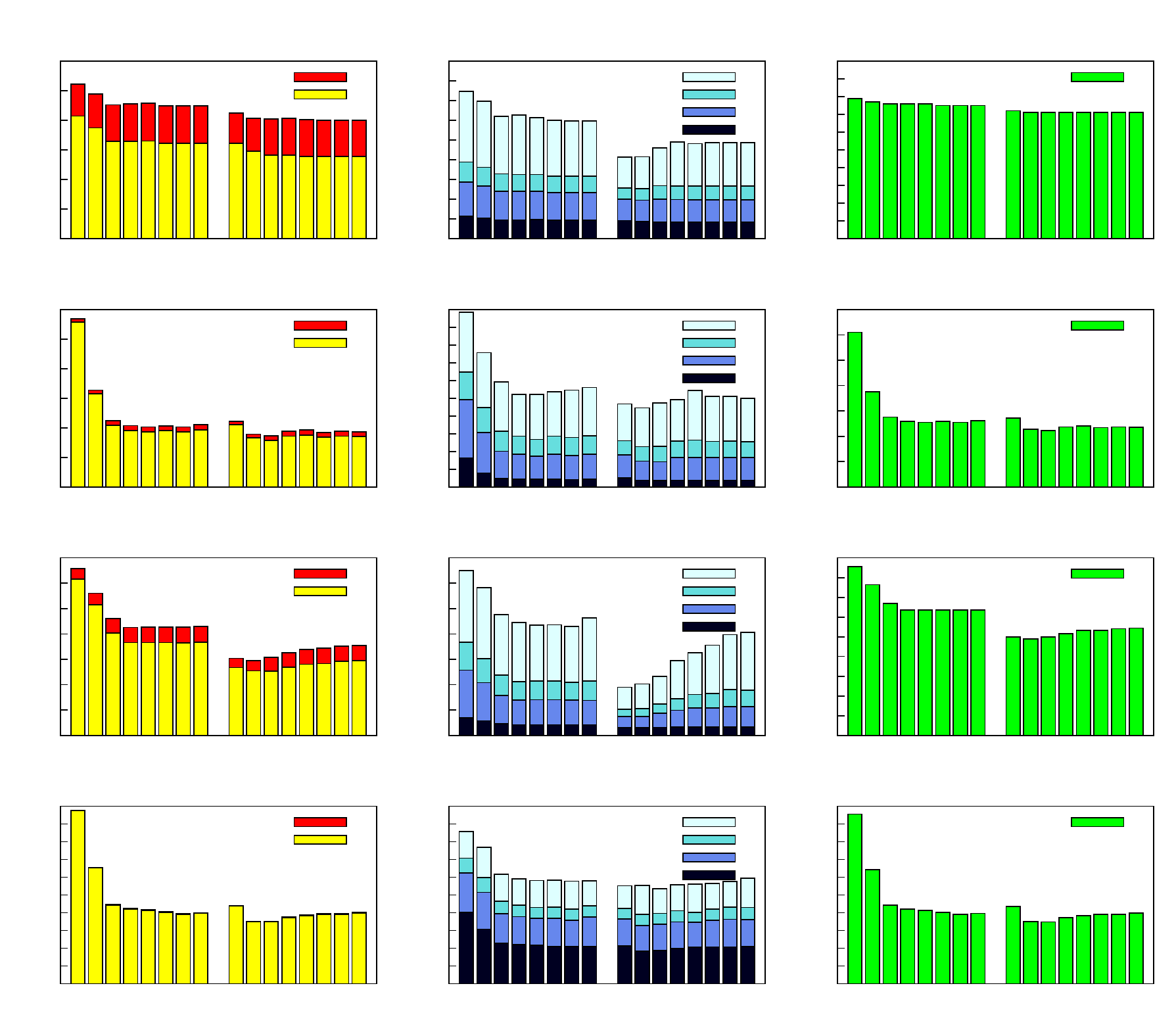}}%
    \gplfronttext
  \end{picture}%
\endgroup

%% file: FIG5-15-1127.tex
% GNUPLOT: LaTeX picture with Postscript
\begingroup
  \makeatletter
  \providecommand\color[2][]{%
    \GenericError{(gnuplot) \space\space\space\@spaces}{%
      Package color not loaded in conjunction with
      terminal option `colourtext'%
    }{See the gnuplot documentation for explanation.%
    }{Either use 'blacktext' in gnuplot or load the package
      color.sty in LaTeX.}%
    \renewcommand\color[2][]{}%
  }%
  \providecommand\includegraphics[2][]{%
    \GenericError{(gnuplot) \space\space\space\@spaces}{%
      Package graphicx or graphics not loaded%
    }{See the gnuplot documentation for explanation.%
    }{The gnuplot epslatex terminal needs graphicx.sty or graphics.sty.}%
    \renewcommand\includegraphics[2][]{}%
  }%
  \providecommand\rotatebox[2]{#2}%
  \@ifundefined{ifGPcolor}{%
    \newif\ifGPcolor
    \GPcolorfalse
  }{}%
  \@ifundefined{ifGPblacktext}{%
    \newif\ifGPblacktext
    \GPblacktexttrue
  }{}%
  % define a \g@addto@macro without @ in the name:
  \let\gplgaddtomacro\g@addto@macro
  % define empty templates for all commands taking text:
  \gdef\gplbacktext{}%
  \gdef\gplfronttext{}%
  \makeatother
  \ifGPblacktext
    % no textcolor at all
    \def\colorrgb#1{}%
    \def\colorgray#1{}%
  \else
    % gray or color?
    \ifGPcolor
      \def\colorrgb#1{\color[rgb]{#1}}%
      \def\colorgray#1{\color[gray]{#1}}%
      \expandafter\def\csname LTw\endcsname{\color{white}}%
      \expandafter\def\csname LTb\endcsname{\color{black}}%
      \expandafter\def\csname LTa\endcsname{\color{black}}%
      \expandafter\def\csname LT0\endcsname{\color[rgb]{1,0,0}}%
      \expandafter\def\csname LT1\endcsname{\color[rgb]{0,1,0}}%
      \expandafter\def\csname LT2\endcsname{\color[rgb]{0,0,1}}%
      \expandafter\def\csname LT3\endcsname{\color[rgb]{1,0,1}}%
      \expandafter\def\csname LT4\endcsname{\color[rgb]{0,1,1}}%
      \expandafter\def\csname LT5\endcsname{\color[rgb]{1,1,0}}%
      \expandafter\def\csname LT6\endcsname{\color[rgb]{0,0,0}}%
      \expandafter\def\csname LT7\endcsname{\color[rgb]{1,0.3,0}}%
      \expandafter\def\csname LT8\endcsname{\color[rgb]{0.5,0.5,0.5}}%
    \else
      % gray
      \def\colorrgb#1{\color{black}}%
      \def\colorgray#1{\color[gray]{#1}}%
      \expandafter\def\csname LTw\endcsname{\color{white}}%
      \expandafter\def\csname LTb\endcsname{\color{black}}%
      \expandafter\def\csname LTa\endcsname{\color{black}}%
      \expandafter\def\csname LT0\endcsname{\color{black}}%
      \expandafter\def\csname LT1\endcsname{\color{black}}%
      \expandafter\def\csname LT2\endcsname{\color{black}}%
      \expandafter\def\csname LT3\endcsname{\color{black}}%
      \expandafter\def\csname LT4\endcsname{\color{black}}%
      \expandafter\def\csname LT5\endcsname{\color{black}}%
      \expandafter\def\csname LT6\endcsname{\color{black}}%
      \expandafter\def\csname LT7\endcsname{\color{black}}%
      \expandafter\def\csname LT8\endcsname{\color{black}}%
    \fi
  \fi
  \setlength{\unitlength}{0.0500bp}%
  \begin{picture}(10318.00,4534.00)%
    \gplgaddtomacro\gplbacktext{%
      \csname LTb\endcsname%
      \put(396,2616){\makebox(0,0)[r]{\strut{} 0}}%
      \put(396,3005){\makebox(0,0)[r]{\strut{} 0.05}}%
      \put(396,3393){\makebox(0,0)[r]{\strut{} 0.1}}%
      \put(396,3782){\makebox(0,0)[r]{\strut{} 0.15}}%
      \put(396,4170){\makebox(0,0)[r]{\strut{} 0.2}}%
      \put(682,2517){\makebox(0,0){\strut{}0}}%
      \put(836,2517){\makebox(0,0){\strut{}1}}%
      \put(990,2517){\makebox(0,0){\strut{}2}}%
      \put(1144,2517){\makebox(0,0){\strut{}3}}%
      \put(1298,2517){\makebox(0,0){\strut{}4}}%
      \put(1452,2517){\makebox(0,0){\strut{}5}}%
      \put(1606,2517){\makebox(0,0){\strut{}6}}%
      \put(1760,2517){\makebox(0,0){\strut{}7}}%
      \put(2069,2517){\makebox(0,0){\strut{}0}}%
      \put(2223,2517){\makebox(0,0){\strut{}1}}%
      \put(2377,2517){\makebox(0,0){\strut{}2}}%
      \put(2531,2517){\makebox(0,0){\strut{}3}}%
      \put(2685,2517){\makebox(0,0){\strut{}4}}%
      \put(2839,2517){\makebox(0,0){\strut{}5}}%
      \put(2993,2517){\makebox(0,0){\strut{}6}}%
      \put(3147,2517){\makebox(0,0){\strut{}7}}%
      \put(22,3393){\rotatebox{-270}{\makebox(0,0){\strut{}Mean latency ($\overline{d}$) [$\unit[]{ms}$]}}}%
      \put(1914,2198){\makebox(0,0){\strut{} }}%
    }%
    \gplgaddtomacro\gplfronttext{%
      \csname LTb\endcsname%
      \put(2446,3722){\makebox(0,0)[r]{\strut{}$\overline{d}_Q$}}%
      \csname LTb\endcsname%
      \put(2446,3876){\makebox(0,0)[r]{\strut{}$\overline{d}_T$}}%
      \csname LTb\endcsname%
      \put(2446,4030){\makebox(0,0)[r]{\strut{}$\overline{d}_R$}}%
      \csname LTb\endcsname%
      \put(2608,2363){\makebox(0,0){\strut{}RDA/R}}%
      \put(1221,2363){\makebox(0,0){\strut{}RDA/Q}}%
    }%
    \gplgaddtomacro\gplbacktext{%
      \csname LTb\endcsname%
      \put(396,440){\makebox(0,0)[r]{\strut{} 0}}%
      \put(396,570){\makebox(0,0)[r]{\strut{} 0.1}}%
      \put(396,699){\makebox(0,0)[r]{\strut{} 0.2}}%
      \put(396,829){\makebox(0,0)[r]{\strut{} 0.3}}%
      \put(396,958){\makebox(0,0)[r]{\strut{} 0.4}}%
      \put(396,1088){\makebox(0,0)[r]{\strut{} 0.5}}%
      \put(396,1218){\makebox(0,0)[r]{\strut{} 0.6}}%
      \put(396,1347){\makebox(0,0)[r]{\strut{} 0.7}}%
      \put(396,1477){\makebox(0,0)[r]{\strut{} 0.8}}%
      \put(396,1606){\makebox(0,0)[r]{\strut{} 0.9}}%
      \put(396,1736){\makebox(0,0)[r]{\strut{} 1}}%
      \put(396,1865){\makebox(0,0)[r]{\strut{} 1.1}}%
      \put(396,1995){\makebox(0,0)[r]{\strut{} 1.2}}%
      \put(682,341){\makebox(0,0){\strut{}0}}%
      \put(836,341){\makebox(0,0){\strut{}1}}%
      \put(990,341){\makebox(0,0){\strut{}2}}%
      \put(1144,341){\makebox(0,0){\strut{}3}}%
      \put(1298,341){\makebox(0,0){\strut{}4}}%
      \put(1452,341){\makebox(0,0){\strut{}5}}%
      \put(1606,341){\makebox(0,0){\strut{}6}}%
      \put(1760,341){\makebox(0,0){\strut{}7}}%
      \put(2069,341){\makebox(0,0){\strut{}0}}%
      \put(2223,341){\makebox(0,0){\strut{}1}}%
      \put(2377,341){\makebox(0,0){\strut{}2}}%
      \put(2531,341){\makebox(0,0){\strut{}3}}%
      \put(2685,341){\makebox(0,0){\strut{}4}}%
      \put(2839,341){\makebox(0,0){\strut{}5}}%
      \put(2993,341){\makebox(0,0){\strut{}6}}%
      \put(3147,341){\makebox(0,0){\strut{}7}}%
      \put(88,1217){\rotatebox{-270}{\makebox(0,0){\strut{}Mean latency ($\overline{d}$) [$\unit[]{ms}$]}}}%
      \put(1914,22){\makebox(0,0){\strut{} }}%
    }%
    \gplgaddtomacro\gplfronttext{%
      \csname LTb\endcsname%
      \put(2446,1547){\makebox(0,0)[r]{\strut{}$\overline{d}_Q$}}%
      \csname LTb\endcsname%
      \put(2446,1701){\makebox(0,0)[r]{\strut{}$\overline{d}_T$}}%
      \csname LTb\endcsname%
      \put(2446,1855){\makebox(0,0)[r]{\strut{}$\overline{d}_R$}}%
      \csname LTb\endcsname%
      \put(2608,187){\makebox(0,0){\strut{}RDA/R}}%
      \put(1221,187){\makebox(0,0){\strut{}RDA/Q}}%
    }%
    \gplgaddtomacro\gplbacktext{%
      \csname LTb\endcsname%
      \put(3800,2616){\makebox(0,0)[r]{\strut{} 0}}%
      \put(3800,2875){\makebox(0,0)[r]{\strut{} 1}}%
      \put(3800,3134){\makebox(0,0)[r]{\strut{} 2}}%
      \put(3800,3393){\makebox(0,0)[r]{\strut{} 3}}%
      \put(3800,3652){\makebox(0,0)[r]{\strut{} 4}}%
      \put(3800,3911){\makebox(0,0)[r]{\strut{} 5}}%
      \put(3800,4170){\makebox(0,0)[r]{\strut{} 6}}%
      \put(4086,2517){\makebox(0,0){\strut{}0}}%
      \put(4240,2517){\makebox(0,0){\strut{}1}}%
      \put(4394,2517){\makebox(0,0){\strut{}2}}%
      \put(4548,2517){\makebox(0,0){\strut{}3}}%
      \put(4703,2517){\makebox(0,0){\strut{}4}}%
      \put(4857,2517){\makebox(0,0){\strut{}5}}%
      \put(5011,2517){\makebox(0,0){\strut{}6}}%
      \put(5165,2517){\makebox(0,0){\strut{}7}}%
      \put(5473,2517){\makebox(0,0){\strut{}0}}%
      \put(5627,2517){\makebox(0,0){\strut{}1}}%
      \put(5781,2517){\makebox(0,0){\strut{}2}}%
      \put(5935,2517){\makebox(0,0){\strut{}3}}%
      \put(6090,2517){\makebox(0,0){\strut{}4}}%
      \put(6244,2517){\makebox(0,0){\strut{}5}}%
      \put(6398,2517){\makebox(0,0){\strut{}6}}%
      \put(6552,2517){\makebox(0,0){\strut{}7}}%
      \put(3624,3393){\rotatebox{-270}{\makebox(0,0){\strut{}Latency ($d$) [$\unit[]{ms}$]}}}%
      \put(5319,2198){\makebox(0,0){\strut{} }}%
      \put(5319,4302){\makebox(0,0){\strut{}\normalsize Benign environment $E_B$ --- \emph{c}(1)}}%
    }%
    \gplgaddtomacro\gplfronttext{%
      \csname LTb\endcsname%
      \put(5851,3568){\makebox(0,0)[r]{\strut{}$d_{p95}$}}%
      \csname LTb\endcsname%
      \put(5851,3722){\makebox(0,0)[r]{\strut{}$d_{p99}$}}%
      \csname LTb\endcsname%
      \put(5851,3876){\makebox(0,0)[r]{\strut{}$d_{p99.5}$}}%
      \csname LTb\endcsname%
      \put(5851,4030){\makebox(0,0)[r]{\strut{}$d_{p99.9}$}}%
      \csname LTb\endcsname%
      \put(6013,2363){\makebox(0,0){\strut{}RDA/R}}%
      \put(4626,2363){\makebox(0,0){\strut{}RDA/Q}}%
    }%
    \gplgaddtomacro\gplbacktext{%
      \csname LTb\endcsname%
      \put(3800,440){\makebox(0,0)[r]{\strut{} 0}}%
      \put(3800,699){\makebox(0,0)[r]{\strut{} 10}}%
      \put(3800,958){\makebox(0,0)[r]{\strut{} 20}}%
      \put(3800,1218){\makebox(0,0)[r]{\strut{} 30}}%
      \put(3800,1477){\makebox(0,0)[r]{\strut{} 40}}%
      \put(3800,1736){\makebox(0,0)[r]{\strut{} 50}}%
      \put(3800,1995){\makebox(0,0)[r]{\strut{} 60}}%
      \put(4086,341){\makebox(0,0){\strut{}0}}%
      \put(4240,341){\makebox(0,0){\strut{}1}}%
      \put(4394,341){\makebox(0,0){\strut{}2}}%
      \put(4548,341){\makebox(0,0){\strut{}3}}%
      \put(4703,341){\makebox(0,0){\strut{}4}}%
      \put(4857,341){\makebox(0,0){\strut{}5}}%
      \put(5011,341){\makebox(0,0){\strut{}6}}%
      \put(5165,341){\makebox(0,0){\strut{}7}}%
      \put(5473,341){\makebox(0,0){\strut{}0}}%
      \put(5627,341){\makebox(0,0){\strut{}1}}%
      \put(5781,341){\makebox(0,0){\strut{}2}}%
      \put(5935,341){\makebox(0,0){\strut{}3}}%
      \put(6090,341){\makebox(0,0){\strut{}4}}%
      \put(6244,341){\makebox(0,0){\strut{}5}}%
      \put(6398,341){\makebox(0,0){\strut{}6}}%
      \put(6552,341){\makebox(0,0){\strut{}7}}%
      \put(3558,1217){\rotatebox{-270}{\makebox(0,0){\strut{}Latency ($d$) [$\unit[]{ms}$]}}}%
      \put(5319,22){\makebox(0,0){\strut{} }}%
      \put(5319,2127){\makebox(0,0){\strut{}\normalsize Hostile environment $E_H$ --- \emph{c}(1)}}%
    }%
    \gplgaddtomacro\gplfronttext{%
      \csname LTb\endcsname%
      \put(5851,1393){\makebox(0,0)[r]{\strut{}$d_{p95}$}}%
      \csname LTb\endcsname%
      \put(5851,1547){\makebox(0,0)[r]{\strut{}$d_{p99}$}}%
      \csname LTb\endcsname%
      \put(5851,1701){\makebox(0,0)[r]{\strut{}$d_{p99.5}$}}%
      \csname LTb\endcsname%
      \put(5851,1855){\makebox(0,0)[r]{\strut{}$d_{p99.9}$}}%
      \csname LTb\endcsname%
      \put(6013,187){\makebox(0,0){\strut{}RDA/R}}%
      \put(4626,187){\makebox(0,0){\strut{}RDA/Q}}%
    }%
    \gplgaddtomacro\gplbacktext{%
      \csname LTb\endcsname%
      \put(7205,2616){\makebox(0,0)[r]{\strut{} 0}}%
      \put(7205,2875){\makebox(0,0)[r]{\strut{} 0.1}}%
      \put(7205,3134){\makebox(0,0)[r]{\strut{} 0.2}}%
      \put(7205,3393){\makebox(0,0)[r]{\strut{} 0.3}}%
      \put(7205,3652){\makebox(0,0)[r]{\strut{} 0.4}}%
      \put(7205,3911){\makebox(0,0)[r]{\strut{} 0.5}}%
      \put(7205,4170){\makebox(0,0)[r]{\strut{} 0.6}}%
      \put(7491,2517){\makebox(0,0){\strut{}0}}%
      \put(7645,2517){\makebox(0,0){\strut{}1}}%
      \put(7799,2517){\makebox(0,0){\strut{}2}}%
      \put(7953,2517){\makebox(0,0){\strut{}3}}%
      \put(8107,2517){\makebox(0,0){\strut{}4}}%
      \put(8261,2517){\makebox(0,0){\strut{}5}}%
      \put(8415,2517){\makebox(0,0){\strut{}6}}%
      \put(8569,2517){\makebox(0,0){\strut{}7}}%
      \put(8878,2517){\makebox(0,0){\strut{}0}}%
      \put(9032,2517){\makebox(0,0){\strut{}1}}%
      \put(9186,2517){\makebox(0,0){\strut{}2}}%
      \put(9340,2517){\makebox(0,0){\strut{}3}}%
      \put(9494,2517){\makebox(0,0){\strut{}4}}%
      \put(9648,2517){\makebox(0,0){\strut{}5}}%
      \put(9802,2517){\makebox(0,0){\strut{}6}}%
      \put(9956,2517){\makebox(0,0){\strut{}7}}%
      \put(6897,3393){\rotatebox{-270}{\makebox(0,0){\strut{}Mean queue size ($\overline{q}$) [$\unit[]{frames}$]}}}%
      \put(8723,2198){\makebox(0,0){\strut{} }}%
    }%
    \gplgaddtomacro\gplfronttext{%
      \csname LTb\endcsname%
      \put(9255,4030){\makebox(0,0)[r]{\strut{}$\overline{q}$}}%
      \csname LTb\endcsname%
      \put(9417,2363){\makebox(0,0){\strut{}RDA/R}}%
      \put(8030,2363){\makebox(0,0){\strut{}RDA/Q}}%
    }%
    \gplgaddtomacro\gplbacktext{%
      \csname LTb\endcsname%
      \put(7205,440){\makebox(0,0)[r]{\strut{} 0}}%
      \put(7205,699){\makebox(0,0)[r]{\strut{} 0.25}}%
      \put(7205,958){\makebox(0,0)[r]{\strut{} 0.5}}%
      \put(7205,1218){\makebox(0,0)[r]{\strut{} 0.75}}%
      \put(7205,1477){\makebox(0,0)[r]{\strut{} 1}}%
      \put(7205,1736){\makebox(0,0)[r]{\strut{} 1.25}}%
      \put(7205,1995){\makebox(0,0)[r]{\strut{} 1.5}}%
      \put(7491,341){\makebox(0,0){\strut{}0}}%
      \put(7645,341){\makebox(0,0){\strut{}1}}%
      \put(7799,341){\makebox(0,0){\strut{}2}}%
      \put(7953,341){\makebox(0,0){\strut{}3}}%
      \put(8107,341){\makebox(0,0){\strut{}4}}%
      \put(8261,341){\makebox(0,0){\strut{}5}}%
      \put(8415,341){\makebox(0,0){\strut{}6}}%
      \put(8569,341){\makebox(0,0){\strut{}7}}%
      \put(8878,341){\makebox(0,0){\strut{}0}}%
      \put(9032,341){\makebox(0,0){\strut{}1}}%
      \put(9186,341){\makebox(0,0){\strut{}2}}%
      \put(9340,341){\makebox(0,0){\strut{}3}}%
      \put(9494,341){\makebox(0,0){\strut{}4}}%
      \put(9648,341){\makebox(0,0){\strut{}5}}%
      \put(9802,341){\makebox(0,0){\strut{}6}}%
      \put(9956,341){\makebox(0,0){\strut{}7}}%
      \put(6870,1217){\rotatebox{-270}{\makebox(0,0){\strut{}Mean queue size ($\overline{q}$) [$\unit[]{frames}$]}}}%
      \put(8723,22){\makebox(0,0){\strut{} }}%
    }%
    \gplgaddtomacro\gplfronttext{%
      \csname LTb\endcsname%
      \put(9255,1855){\makebox(0,0)[r]{\strut{}$\overline{q}$}}%
      \csname LTb\endcsname%
      \put(9417,187){\makebox(0,0){\strut{}RDA/R}}%
      \put(8030,187){\makebox(0,0){\strut{}RDA/Q}}%
    }%
    \gplbacktext
    \put(0,0){\includegraphics{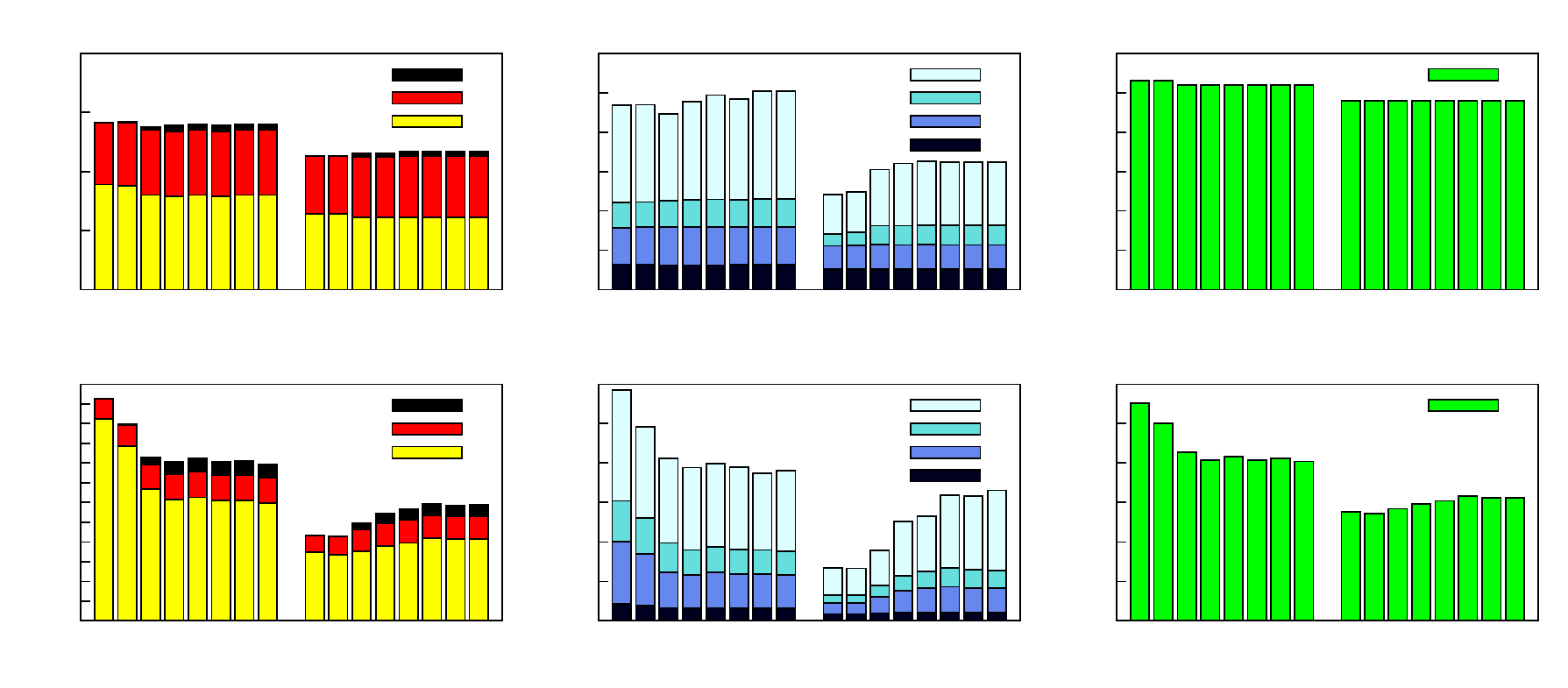}}%
    \gplfronttext
  \end{picture}%
\endgroup